\documentclass[
 amsmath,amssymb,
 aps,
 pra,
twocolumn
]{revtex4-2}

\usepackage{graphicx}
\usepackage{dcolumn}
\usepackage{bm}
\usepackage{bbm}
\usepackage{comment}
\usepackage{physics}
\usepackage[dvipsnames]{xcolor}
\usepackage{tikz}
\usetikzlibrary{quantikz}
\usepackage[ruled,vlined]{algorithm2e}
\usepackage{graphicx}
\usepackage{subfigure}
\newtheorem{theorem}{Theorem}

\newcommand{\val}{\text{val}}

\newcommand{\J}{\text{junk}}
\newcommand{\svec}{\text{vec}}
\usepackage[breaklinks=true]{hyperref}
\hypersetup{
  colorlinks   = true, 
  urlcolor     = blue, 
  linkcolor    = blue, 
  citecolor   = magenta 
}
\usepackage{soul} 
\newtheorem{lemma}{Lemma}

\begin{document}

\preprint{APS/123-QED}

\title{Quantum diffusion map for nonlinear dimensionality reduction }

\author{Apimuk Sornsaeng}
\author{Ninnat Dangniam}
\author{Pantita Palittapongarnpim}
\email[Correspondence to ]{panpalitta@gmail.com}
\author{Thiparat Chotibut}%
 \email[Correspondence to ]{thiparatc@gmail.com, thiparat.c@chula.ac.th}
\affiliation{%
 Chula Intelligent and Complex Systems, Department of Physics, Faculty of Science, Chulalongkorn University, Bangkok, Thailand, 10330
}

\date{\today}
\begin{abstract}
 
Inspired by random walk on graphs, diffusion map (DM) is a class of unsupervised machine learning that offers automatic identification of low-dimensional data structure hidden in a high-dimensional dataset. In recent years, among its many applications, DM has been successfully applied to discover relevant order parameters in many-body systems, enabling automatic classification of quantum phases of matter.
However, classical DM algorithm is computationally prohibitive for a large dataset, and any reduction of the time complexity would be desirable. With a quantum computational speedup in mind, we propose a quantum algorithm for DM, termed {\it quantum diffusion map} (qDM). Our qDM takes as an input $N$ classical data vectors, performs an eigen-decomposition of the Markov transition matrix in time $O(\log^3 N)$, and classically constructs the diffusion map via the readout (tomography) of the eigenvectors, giving a total {expected runtime proportional to $N^2 \text{polylog}\, N$}.
Lastly, quantum subroutines in qDM for constructing a Markov transition matrix, and for analyzing its spectral properties can also be useful for other random walk-based algorithms.  
\end{abstract}

\maketitle


\section{INTRODUCTION}\label{sec:level1}

Discovering statistical structure in high-dimensional data is essential for data-driven science and engineering. Advances in unsupervised machine learning offer a plethora of alternatives to automatically search for low-dimensional structure of data lying in a high-dimensional space. Many of these approaches involve dimensionality reduction. Classic approach is the principal component analysis (PCA), which projects high-dimensional data onto a low-dimensional linear space spanned by a set of orthonormal bases, whose directions capture significant data variations. More compelling approach enables automatic searches for a low-dimensional data manifold embedded in a high-dimensional space. Well-known manifold learning algorithms include Isomap \cite{Tenenbaum2319}, Laplacian eigenmaps \cite{BelkinLapEig}, uniform manifold approximation and projection (UMAP) \cite{McInnes2018}, nonlinear PCA \cite{Scholz_nonlinpca}, t-distributed stochastic neighbor embedding (t-SNE) \cite{tSNE_JMLR}, and diffusion map (DM) \cite{coifman2006diffusion,lafon2004diffusion, DM_NIPS2006}, which is the focus of this paper. Inspired by appealing features of random walk on graphs, DM and its variants have received increasing attention for data visualization in bioinformatics \cite{PHATE_nat2019, multiPHATE_2019}. DM has also recently been sucessfully applied to provide automatic classification of topological phases of matter, and offers automatic identification of quantum phase transitions in many-body systems \cite{scheurer_natphys_2019, Chen_PRL_TopoPhonon2019,Gong_PRL_2020, Kerr_PRE_2021,nori_PRB_2021, wang_PRR_2021}.

Most dimensionality reduction methods require the computation of singular value decomposition (SVD) of a matrix constructed from a collection of high-dimensional data points. For a matrix of size, say, $N \times d$, where $N$ is the number of data points and $d$ is the dimensionality of each data vector (assumed to be smaller than $N$), the computational cost of SVD typically grows with the number of data points as $O(N^3)$. Thus, classical dimensionality reduction can be computationally prohibitive for a large data sample. 
However, under moderate assumptions of accessibility to certain features of full-scale quantum computers, matrix exponentiation-based quantum algorithms have been proposed to perform SVD more efficiently \cite{lloyd_PRA_QSVD}. In particular, assuming efficient encoding of classical data to quantum information and accessibility to appropriate quantum RAM, quantum singular value decomposition (qSVD) algorithm's runtime for many non-sparse low-rank Hermitian matrices is $O(\text{polylog}\,N)$, which is exponentially faster than the classical counterpart and can be extended to non-square matrices. A classic quantum algorithm for dimensionality reduction with quantum computational speedup is quantum principal component analysis (qPCA), which exploits matrix exponentiation tricks as in qSVD \cite{lloyd2014quantum}. More recently, quantum algorithms with quantum advantage for nonlinear dimensionality reduction with nonlinear kernel~\cite{qKPCA} and cluster identification based on spectral graph theory have also been proposed \cite{dewolf_ieee_2020,Kerenidis_PRB2021,q-means_NEURIPS2019}.

With a quantum computational speedup for dimensionality reduction in mind, we propose a quantum algorithm for unsupervised manifold learning called quantum diffusion map (qDM). Under mild assumptions of appropriate oracles, qDM has {an expected} runtime of 
roughly {$\kappa_D^{0.625} N^2 \mathrm{polylog}\,N$}
where $N$ is the number of data points and $\kappa_D$ is the condition number of the degree matrix (see Theorem \ref{thm:main} and Lemma \ref{lemma: classical_coupon} for the precise statement), as opposed to $O(N^3)$ in classical DM. 
Although $\kappa_D$ depends strongly on the data structure and can take the value $N$ in the worst case, such worst-case is highly atypical for well structured dataset, in which $\kappa_D = O(1)$. Without the final readout step, our qDM algorithm prepares all necessary components for constructing diffusion map in {$O(\kappa_D^{0.625}\log^3N)$} time.

Although the backbone of DM is a graph-based dimensionality reduction method, the procedure is different from other spectral graph methods. Namely, rather than working with the data-induced graph Laplacian as in Laplacian eigenmaps or in spectral clustering, DM involves Markov transition matrix that defines random walks on a data-induced graph. Therefore, recipes for qDM dimensionality reduction are different from the recently proposed quantum spectral clustering \cite{dewolf_ieee_2020,Kerenidis_PRB2021}{, mainly in the efficient construction of the degree matrix, which was mentioned as not efficiently accessible~\cite{Kerenidis_PRB2021}. Our algorithm also exploits the coherent state encoding scheme, initially proposed in Ref.~\cite{chatterjee2016generalized}, to provide an exact quantum encoding of classical Gaussian kernel. This is in contrast to the quantum kernel principle component analysis (qKPCA) \cite{qKPCA} for dimensionality reduction, which only encodes nonlinear kernel approximately.}

The paper is organized as follows. Section \ref{sec: cDM} provides necessary ingredients to understand classical diffusion map and its runtime. Section \ref{sec: roadmap} introduces quantum subroutines for qDM, which is then discussed in details in Section \ref{sec: transition_matrix} and \ref{sec: QPE2}. qDM runtime complexity appears in Section \ref{sec:complexity_quantum}. We conclude with the discussion and outlook in Section \ref{sec: conclude}. Additional details on the algorithmic complexity and the subroutines for qDM are provided in Appendix \ref{sec:QMAT} and \ref{sec: complexity_ana}. 

\section{Classical diffusion map for dimensionality reduction}\label{sec: cDM}

\begin{figure*}
    \centering     
    \subfigure[]{\label{fig:a}\includegraphics[scale=0.4]{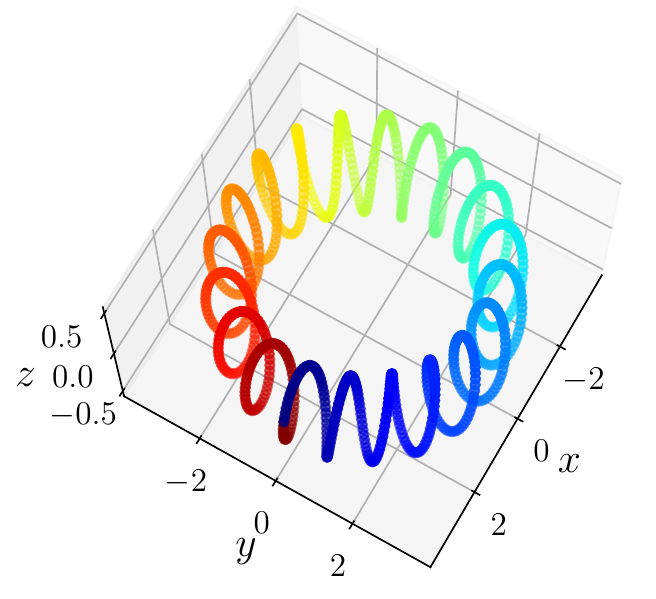}}
    \subfigure[]{\label{fig:b}\includegraphics[scale=0.4]{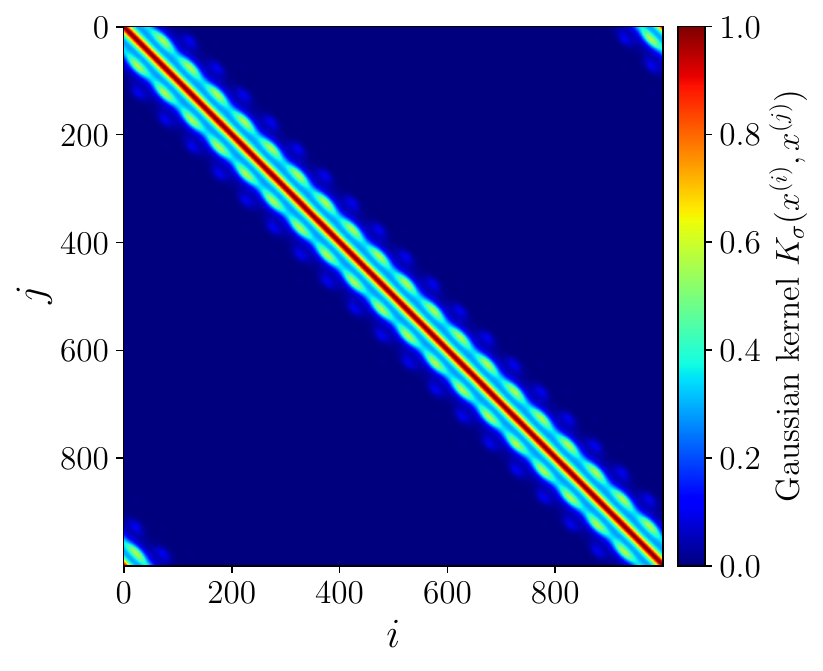}}
    \subfigure[]{\label{fig:c}\includegraphics[scale=0.4]{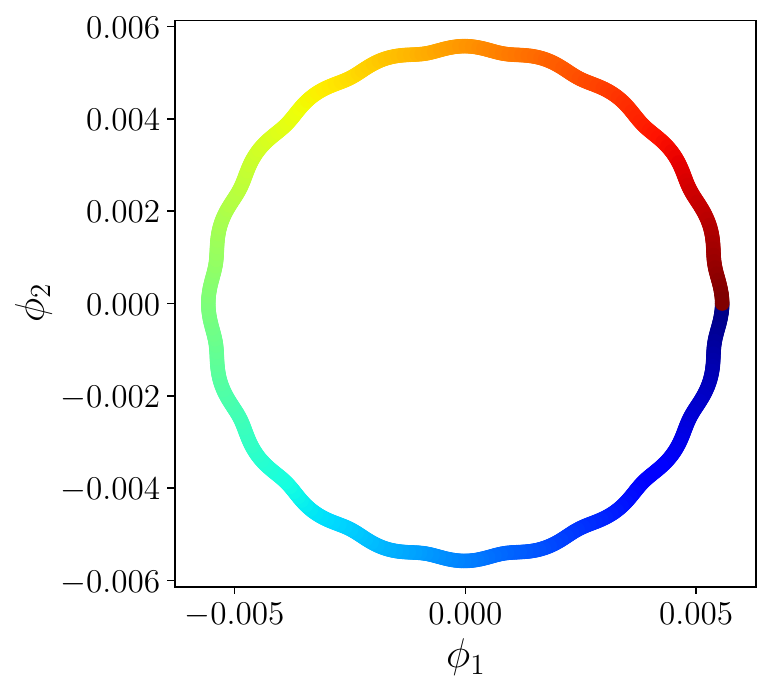}}
    \subfigure[]{\label{fig:d}\includegraphics[scale=0.4]{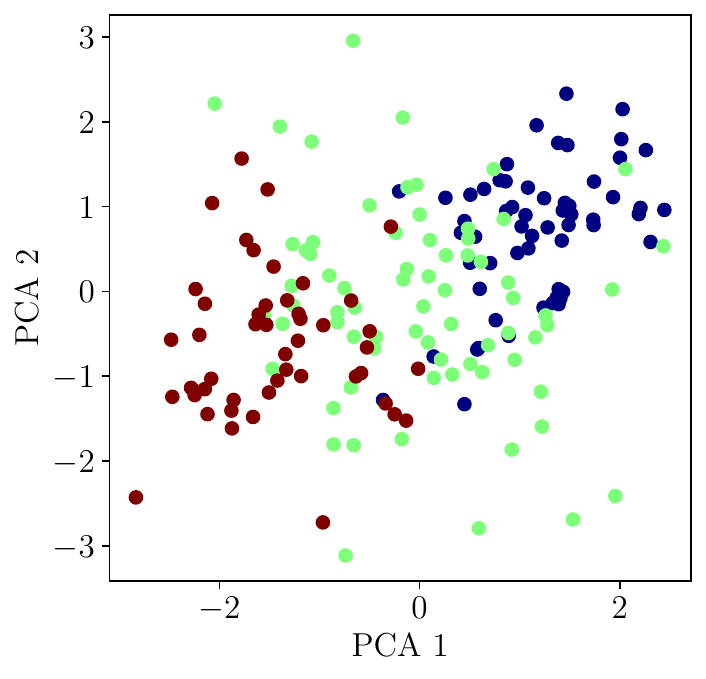}}
    \subfigure[]{\label{fig:e}\includegraphics[scale=0.4]{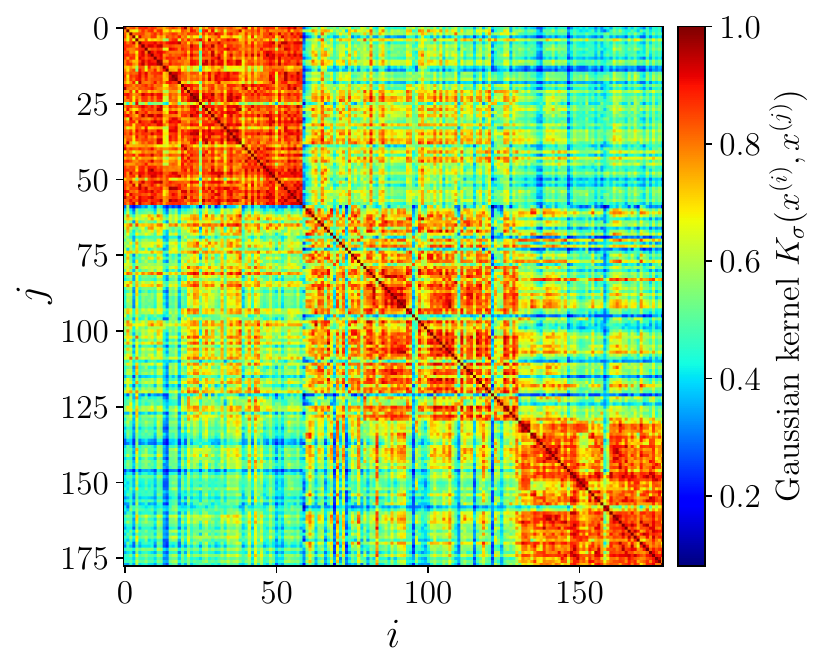}}
    \subfigure[]{\label{fig:f}\includegraphics[scale=0.4]{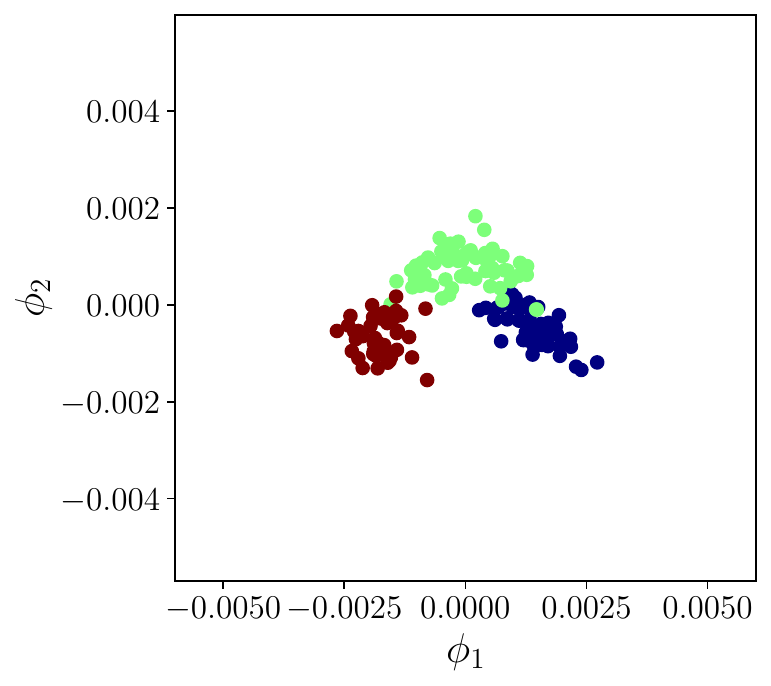}}
\caption{The application of classical DM for nonlinear dimensionality reduction: (top) to identify the proximity structure of data points distributed as a toroidal helix in 3 dimensions, and (bottom) to reveal the proximity structure of 13 chemical components' concentrations of 178 wines ($d=13, N=178$) derived from 3 different cultivars grown in the same region in Italy, taken from the WINE dataset in \cite{Dua:2019}. In a toroidal helix data (top left), color variation signifies difference in the geodesic distance between points on a one-dimensional manifold. The Gaussian kernel with $\sigma=1$ of \eqref{eq: kernel} shows the neighborhood connectivity has a local, periodic structure (top middle). The dimensionality reduction (top right) into the first two components ($m=2$) of the DM in \eqref{eq: DM} reveals a one-dimensional structure with an appropriate notion of geodesic distance (see color variations). In WINE dataset, wines grown from 3 different cultivars are labeled in 3 different colors (red, green, blue). Projecting the 13 data features (chemical concentrations) of all wines into the first two principal components using PCA does not suffice to distinguish the 3 types (bottom left). However, with the Gaussian kernel with $\sigma=50$ (bottom middle), DM with two bases ($m=2$) can reveal quite a clear distinction between types (bottom right). In addition, the proximity structure also suggests that the green type has its chemical components closer to those of the other two types. In both dataset, we use $t=1$ in the DM. Thus, random walk on data-induced graphs can reveal salient low-dimensional data structure within just a one-step walk, provided $\sigma$ is chosen appropriately. 
}
\label{fig: DM_ex}
\end{figure*}

Given a set of $N$ data vectors $X \equiv \qty{\boldsymbol x^{(i)} }_{i=0}^{N-1}$, in which each data vector has $d$ features (i.e. $\boldsymbol x^{(i)} \in \mathbb{R}^d$), unsupervised machine learning seeks to identify the structure of data distribution in a high-dimensional ($d\gg1$) space. When the data distribution has a highly nonlinear structure, classic linear approach such as PCA fails. Rather than using mere Euclidean distance as a similarity measure between data points, manifold learning approach assigns the connectivity among data points in their neighborhood as a proxy for proximity between points; e.g., points on a toroidal helix embedded in 3 dimensions with equal Euclidean distance can have different geodesic distances on a manifold (different similarity), see Fig. \ref{fig: DM_ex} (top). This approach allows one to perform a nonlinear dimensionality reduction by assigning an appropriate function that maps a high-dimensional data point into a relevant lower-dimensional Euclidean space embedding, while encapsulating the notion of proximity in high dimensions based on neighborhood connectivity.

Neighborhood connectivity leads to the development of graph-based manifold learning approach, where one can assign a vertex $i$ to a data vector $\boldsymbol x^{(i)}$ and assign an edge between a pair of data points that are considered to be neighbors. Isomap \cite{Tenenbaum2319} and Laplacian eigenmaps \cite{BelkinLapEig} are among the first graph-based manifold learning algorithms, where relevant low-dimensional data embedding can be extracted from eigen-decomposition of data-induced graph Laplacian matrix. 

\subsection{Diffusion map in a nutshell\label{subsec: classical_DM}}
In diffusion map (DM), the similarity matrix between a pair of data vectors, or equivalently the weighted edge between a pair of graph vertices, is often taken to be a Gaussian kernel:
\begin{equation}\label{eq: kernel}
    W_{ij}\equiv K_\sigma\qty(\vb*{x}^{(i)},\vb*{x}^{(j)})=\exp\qty(-\frac{\norm{\vb*{x}^{(i)}-\vb*{x}^{(j)}}_2^2}{2\sigma}),
\end{equation}
where the adjustable parameter $\sigma$, called the {\it bandwidth}, sets the scale of neighborhood. Two points whose squared distance exceed this bandwidth contribute exponentially little to the weighted edge, suggesting that these two points are far away from being a neighbor in its original feature space.  Given a graph with weighted edges $W_{ij}$'s, DM assigns a discrete-time random walk on a data-induced graph, where the Markov transition probability from vertex $i$ to $j$ is given by the normalized weighted edge,
\begin{equation}\label{eq: prob_tran}
    P_{ij}=\frac{W_{ij}}{\sum_{j=0}^{N-1}W_{ij}}.
\end{equation}
One can compactly compute the Markov transition matrix $P$ from the (weighted) {\it adjacency matrix} $W$, and the {\it degree matrix} $D\equiv \text{diag}\{d_i\}_{i=0}^{N-1}$ where $d_i\equiv\sum_{j=0}^{N-1}W_{ij}$ by 
\begin{equation}\label{eq: P_construction}
	P=D^{-1}W.
\end{equation}

The notion of proximity based on graph connectivity can be characterized by how fast random walkers on a graph starting at different data points visit each other. One expects that two points that are connected by multiple paths should be near; whereas two points that are sparsely connected should lie far from each other. In fact, it can be shown that \cite{DM_NIPS2006} the proper distance function, called {\it diffusion distance}, between two points $\boldsymbol{x}^{(i)}$ and $\boldsymbol{x}^{(j)}$ after a $t$-step random walk is given by
 \begin{equation}\label{diffdist}
 	\operatorname{Dist}_{t}^{2}\left(\boldsymbol{x}^{(i)}, \boldsymbol{x}^{(j)}\right)=\sum_{k=0}^{N-1} \frac{\left[\left(P^{t}\right)_{i k}-\left(P^{t}\right)_{j k}\right]^{2}}{u_{0}\left(\boldsymbol{x}^{(k)}\right)}, 
 \end{equation}
 where $u_{0}$ is the {\it left} eigenvector with eigenvalue $\lambda_0 = 1$ of $P$ and $u_{0}\left(\boldsymbol{x}^{(k)}\right)$ denotes the $k^{\textrm{th}}$ coordinate of $u_{0}$.
 
Before obtaining a relevant lower-dimensional Euclidean space representation of the original data point $\boldsymbol{x}^{(i)}$, we first note the following important identity. Define the {\it diffusion map} with $m$ bases as
 \begin{equation}    \label{eq: DM}
 \vb*{\phi}_{t,m}\qty(\vb*{x}^{(i)})=\begin{pmatrix}\lambda_{1}^tv_{1}\left(\vb*{x}^{(i)}\right)\\
    \vdots\\\lambda_{m}^tv_{m}\left(\vb*{x}^{(i)}\right)\end{pmatrix},
 \end{equation}
where $v_i$ is the {\it right} eigenvector of $P$ with eigenvalue $\lambda_i$, $v_i\left(\boldsymbol{x}^{(j)}\right)$ denotes the $j^{\textrm{th}}$ coordinate of $v_i$, and the eigenvalues are ordered such that $\lambda_{0}=1 > \lambda_{1} \geq \lambda_{2} \geq \ldots \geq \lambda_{N-1} \geq 0$ assuming no degeneracy in the largest eigenvalue (which is always possible given $\sigma$ is sufficiently large). Note that $\lambda_0 = 1$ is always the eigenvalue of a Markov transition matrix $P$ with a normalized eigenvector $\mqty(1&\cdots&1)^T/N$, corresponding to the stationary uniform distribution over all vertices. Then, the diffusion distance \eqref{diffdist} can be {\it exactly} computed from the Euclidean representation in \eqref{eq: DM} with $m = N-1$ bases \cite{DM_NIPS2006};
\begin{equation}\label{eq: identity}
    \operatorname{Dist}_{t}^{2}\left(\boldsymbol{x}^{(i)}, \boldsymbol{x}^{(j)}\right)=\norm{\vb*{\phi}_{t,N-1}\qty(\vb*{x}^{(i)})-\vb*{\phi}_{t,N-1}\qty(\vb*{x}^{(j)})}_2^2.
\end{equation}
The above equality states that the distance in the diffusion space (based on graph connectivity) is identical to the Euclidean embedding distance (induced by the diffusion map). Thus, the notion of proximity between data points from its graph connectivity can be simply computed from the Euclidean embedding by the diffusion map \eqref{eq: DM}.

What about dimensionality reduction? Since $P$ is a Markov transition matrix, its eigenvalues are $\lambda_{0}=1 \geq \lambda_{1} \geq \lambda_{2} \geq \ldots \geq \lambda_{N-1} \geq 0.$ Bases in \eqref{eq: DM} with low-lying eigenvalues are then exponentially suppressed as $t$-step increases. In the long-time limit, dimensionality reduction thus naturally arises in DM. One may take the number of bases $m$ in \eqref{eq: DM}  corresponding to the number of top eigenvalues of $P$, and still obtain meaningful low-dimensional Euclidean representation of points in the diffusion space. In practice, for the purpose of visualization, taking $m=2 \ll d$ is a drastic dimensionality reduction, yet DM with such small number of bases can yield insights into the approximate proximity of data in the original high-dimensional space, see Fig. \ref{fig: DM_ex}.  Using DM \eqref{eq: DM} to extract low-dimensional data representation that approximately preserves the notion of distance from neighborhood connectivity in the original high-dimensional space is also the first step towards data clustering algorithms. Namely, one may employ standard clustering algorithms, such as k-means algorithm, on the low-dimensional outputs of the DM without suffering from the curse of dimensionality problem. 

\subsection{Time complexity of classical diffusion map\label{subsec: classical complexity}}

Numerical recipes to obtain nonlinear dimensionality reduction or manifold learning in classical DM consists of 4 steps:
\begin{enumerate}
	\item Construct the data-induced similarity matrix (weighted adjacency matrix) $W$ from \eqref{eq: kernel} 
	\item Construct the Markov transition matrix $P$ from \eqref{eq: P_construction}; i.e. $P = D^{-1}W$
	\item Compute the eigen-decomposition of $P$
	\item Construct the diffusion map \eqref{eq: DM}
\end{enumerate}

We now discuss the time complexity for each step as a function of the number of data points $N$. Assuming the Gaussian kernel function~(\ref{eq: kernel}), the weighted adjacency matrix $W$ is symmetric. The number of the elements that the algorithm needs to calculate is then $\sum_{i=1}^N i = (N^2+N)/2$. Therefore, the time complexity of calculating the kernel matrix in step 1 is $O\left(N^2\right)$.

For step 2, the calculation of~$P$ involves normalizing each row of~$W$ as in \eqref{eq: prob_tran}. 
Numerical computation of the normalization factor, which is simply the summation, takes the time $O(N)$. Then normalizing each row costs another $O(N)$.
Applying these procedures over the $N\times N$ matrix thus yields the time cost $O(N^2)+O(N^2)$.
Hence, the time complexity for computing $P$ is $O(N^2)$.

In step 3, the time cost stems from finding $\{v_i\}$ and $\{\lambda_i\}$ of $P$, using singular value decomposition (SVD), which can be applied to any non-squared matrices. This algorithm comprises two steps~\cite{trefethen1997numerical}. The first step is to use Householder transformations to reduce $P$ to a bidiagonal form. Then the {QR} algorithm is applied to find singular values. The time complexity of these two steps combined is $O(N^3)$.

The last step is computing the diffusion map $\bm{\phi}$ according to \eqref{eq: DM}, which involves the multiplication of $\lambda_i^t$ with its corresponding right eigenvector $v_i$.
Although we might not need to perform the multiplication for all $N-1$ right eigenvectors as some $\lambda_i^t$ maybe negligible in the long-time limit, we consider the worst case scenario where all $N-1$ eigenvectors are required. With the length of each vector being $N$, the time cost is $O(N^2-N)$ for computing $\bm{\phi}$ and thus the time complexity for this step is $O(N^2)$.

Combining all the steps above, which is implemented in sequence,  classical DM has a time complexity of $O(N^2)+O(N^2)+O(N^3)+O(N^2)$ leading to an overall time complexity that scales as $O(N^3)$. Note here that the time complexity is primarily dominated by the eigen-decomposition algorithm, which is subjected to change if a different algorithm is chosen in the implementation. In fact, this dominating complexity of $O(N^3)$ from eigen-decomposition urges for the development of quantum algorithms to achieve a quantum computational speedup, which we discuss next.

\section{Quantum diffusion map roadmap\label{sec: roadmap}}

\begin{figure}
    \centering

\tikzset{every picture/.style={line width=0.75pt}} 

\begin{tikzpicture}[x=0.75pt,y=0.75pt,yscale=-1,xscale=1]

\draw    (349.04,46) -- (456.04,46) -- (456.04,92) -- (349.04,92) -- cycle  ;
\draw (402.54,69) node   [align=left] {\begin{minipage}[lt]{70.19pt}\setlength\topsep{0pt}
\begin{center}
Input \\Classical State
\end{center}

\end{minipage}};
\draw    (346.04,139.5) -- (461.04,139.5) -- (461.04,185.5) -- (346.04,185.5) -- cycle  ;
\draw (403.54,162.5) node   [align=left] {\begin{minipage}[lt]{75.3pt}\setlength\topsep{0pt}
\begin{center}
Kernel Operator\\(Sec. \ref{subsec: Kernel})
\end{center}

\end{minipage}};
\draw  [color={rgb, 255:red, 0; green, 0; blue, 0 }  ,draw opacity=1 ][fill={rgb, 255:red, 255; green, 255; blue, 255 }  ,fill opacity=1 ]  (353.04,247.5) -- (455.04,247.5) -- (455.04,293.5) -- (353.04,293.5) -- cycle  ;
\draw (404.04,270.5) node   [align=left] {\begin{minipage}[lt]{66.78pt}\setlength\topsep{0pt}
\begin{center}
Degree Matrix\\(Sec. \ref{subsec: exponential})
\end{center}

\end{minipage}};
\draw  [fill={rgb, 255:red, 255; green, 255; blue, 255 }  ,fill opacity=1 ]  (335.54,333) -- (476.54,333) -- (476.54,400) -- (335.54,400) -- cycle  ;
\draw (406.04,366.5) node   [align=left] {\begin{minipage}[lt]{92.87pt}\setlength\topsep{0pt}
\begin{center}
Inverse Square-root\\Degree Matrix\\(Sec. \ref{subsec:invert})
\end{center}

\end{minipage}};
\draw  [fill={rgb, 255:red, 255; green, 255; blue, 255 }  ,fill opacity=1 ]  (348.04,443) -- (466.04,443) -- (466.04,510) -- (348.04,510) -- cycle  ;
\draw (407.04,476.5) node   [align=left] {\begin{minipage}[lt]{77.73pt}\setlength\topsep{0pt}
\begin{center}
Symmetrized\\Transition Matrix\\(Sec. \ref{subsec:invert})
\end{center}

\end{minipage}};
\draw    (334.04,551.5) -- (482.04,551.5) -- (482.04,597.5) -- (334.04,597.5) -- cycle  ;
\draw (408.04,574.5) node   [align=left] {\begin{minipage}[lt]{97.98pt}\setlength\topsep{0pt}
\begin{center}
Eigen-decomposition\\(Sec. \ref{subsec: QPE3})
\end{center}

\end{minipage}};
\draw  [color={rgb, 255:red, 255; green, 0; blue, 0 }  ,draw opacity=1 ]  (357.71,643.17) -- (456.71,643.17) -- (456.71,668.17) -- (357.71,668.17) -- cycle  ;
\draw (407.21,655.67) node   [align=left] {\begin{minipage}[lt]{64.9pt}\setlength\topsep{0pt}
\begin{center}
Diffusion Map
\end{center}

\end{minipage}};
\draw (466.23,115.5) node  [font=\footnotesize] [align=left] {Coherent state \\encoding (Sec. \ref{subsec: coherent})};
\draw  [draw opacity=0][fill={rgb, 255:red, 255; green, 255; blue, 255 }  ,fill opacity=1 ]  (351.67,300.83) -- (398.67,300.83) -- (398.67,320.83) -- (351.67,320.83) -- cycle  ;
\draw (290,310.83) node [anchor=west] [inner sep=0.75pt]  [font=\footnotesize] [align=left] {\begin{minipage}[lt]{100pt}\setlength\topsep{0pt}
\begin{center}
Matrix inversion\\(Sec. \ref{subsec:invert}) 
\end{center}

\end{minipage}};
\draw  [draw opacity=0][fill={rgb, 255:red, 255; green, 255; blue, 255 }  ,fill opacity=1 ]  (418.67,408.5) -- (465.67,408.5) -- (465.67,428.5) -- (418.67,428.5) -- cycle  ;
\draw (410,420) node [anchor=west] [inner sep=0.75pt]  [font=\footnotesize] [align=left] {\begin{minipage}[lt]{80pt}\setlength\topsep{0pt}
\begin{center}
Matrix multiplication\\(Sec. \ref{subsec: exponential}) 
\end{center}

\end{minipage}};
\draw  [draw opacity=0][fill={rgb, 255:red, 255; green, 255; blue, 255 }  ,fill opacity=1 ]  (421.85,198.67) -- (513.85,198.67) -- (513.85,234.67) -- (421.85,234.67) -- cycle  ;
\draw (450,216.67) node  [font=\footnotesize] [align=left] {\begin{minipage}[lt]{59.86pt}\setlength\topsep{0pt}
\begin{center}
Matrix multiplication\\(Sec. \ref{subsec: exponential}) 
\end{center}

\end{minipage}};
\draw  [draw opacity=0][fill={rgb, 255:red, 255; green, 255; blue, 255 }  ,fill opacity=1 ]  (323.44,517.5) -- (398.44,517.5) -- (398.44,537.5) -- (323.44,537.5) -- cycle  ;
\draw (400,530) node [anchor=east] [inner sep=0.75pt]  [font=\footnotesize] [align=left] {\begin{minipage}[lt]{48.52pt}\setlength\topsep{0pt}
\begin{center}
QPE \\ (Sec. \ref{subsec: QPE3}) 
\end{center}

\end{minipage}};
\draw (466.84,620) node  [font=\footnotesize] [align=left] {Readout (Sec. \ref{subsec: extract}) \ };
\draw  [color={rgb, 255:red, 255; green, 0; blue, 0 }  ,draw opacity=1 ][fill={rgb, 255:red, 255; green, 255; blue, 255 }  ,fill opacity=1 ]  (214.71,443.83) -- (332.71,443.83) -- (332.71,489.83) -- (214.71,489.83) -- cycle  ;
\draw (273.71,466.83) node   [align=left] {\begin{minipage}[lt]{77.73pt}\setlength\topsep{0pt}
\begin{center}
Transition Matrix\\(Sec. \ref{subsec:invert})
\end{center}

\end{minipage}};
\draw    (403.3,92.02) -- (403.3,139.5) ;
\draw [shift={(403.3,139.5)}, rotate = 269.39] [color={rgb, 255:red, 0; green, 0; blue, 0 }  ][line width=0.75]    (10.93,-3.29) .. controls (6.95,-1.4) and (3.31,-0.3) .. (0,0) .. controls (3.31,0.3) and (6.95,1.4) .. (10.93,3.29)   ;
\draw    (403.94,185.51) -- (403.94,247.5) ;
\draw [shift={(403.94,247.5)}, rotate = 269.73] [color={rgb, 255:red, 0; green, 0; blue, 0 }  ][line width=0.75]    (10.93,-3.29) .. controls (6.95,-1.4) and (3.31,-0.3) .. (0,0) .. controls (3.31,0.3) and (6.95,1.4) .. (10.93,3.29)   ;
\draw    (405.34,293.53) -- (405.34,333) ;
\draw [shift={(405.34,333)}, rotate = 268.81] [color={rgb, 255:red, 0; green, 0; blue, 0 }  ][line width=0.75]    (10.93,-3.29) .. controls (6.95,-1.4) and (3.31,-0.3) .. (0,0) .. controls (3.31,0.3) and (6.95,1.4) .. (10.93,3.29)   ;
\draw   (407.00,510.02) -- (407.81,551.5) ;
\draw [shift={(407.81,551.5)}, rotate = 269.42] [color={rgb, 255:red, 0; green, 0; blue, 0 }  ][line width=0.75]    (10.93,-3.29) .. controls (6.95,-1.4) and (3.31,-0.3) .. (0,0) .. controls (3.31,0.3) and (6.95,1.4) .. (10.93,3.29)   ;
\draw    (406.35,400) -- (406.72,441) ;
\draw [shift={(406.74,443)}, rotate = 269.48] [color={rgb, 255:red, 0; green, 0; blue, 0 }  ][line width=0.75]    (10.93,-3.29) .. controls (6.95,-1.4) and (3.31,-0.3) .. (0,0) .. controls (3.31,0.3) and (6.95,1.4) .. (10.93,3.29)   ;
\draw    (434.95,293.5) .. controls (553.21,372.71) and (553.24,458.71) .. (435.05,551.5) ;
\draw [shift={(435.05,551.5)}, rotate = 321.86] [color={rgb, 255:red, 0; green, 0; blue, 0 }  ][line width=0.75]    (10.93,-3.29) .. controls (6.95,-1.4) and (3.31,-0.3) .. (0,0) .. controls (3.31,0.3) and (6.95,1.4) .. (10.93,3.29)   ;
\draw    (377.18,185.5) .. controls (264.62,275.39) and (261.88,361.23) .. (368.97,443) ;
\draw [shift={(368.97,443)}, rotate = 217.36] [color={rgb, 255:red, 0; green, 0; blue, 0 }  ][line width=0.75]    (10.93,-3.29) .. controls (6.95,-1.4) and (3.31,-0.3) .. (0,0) .. controls (3.31,0.3) and (6.95,1.4) .. (10.93,3.29)   ;
\draw    (407.34,597.48) -- (407.34,643.17) ;
\draw [shift={(407.34,643.17)}, rotate = 270.59000000000003] [color={rgb, 255:red, 0; green, 0; blue, 0 }  ][line width=0.75]    (10.93,-3.29) .. controls (6.95,-1.4) and (3.31,-0.3) .. (0,0) .. controls (3.31,0.3) and (6.95,1.4) .. (10.93,3.29)   ;
\draw    (361.86,400) -- (305.64,442.62) ;
\draw [shift={(304.04,443.83)}, rotate = 322.83000000000004] [color={rgb, 255:red, 0; green, 0; blue, 0 }  ][line width=0.75]    (10.93,-3.29) .. controls (6.95,-1.4) and (3.31,-0.3) .. (0,0) .. controls (3.31,0.3) and (6.95,1.4) .. (10.93,3.29)   ;
\draw    (376.9,185.5) .. controls (294.68,249.92) and (259.58,335.49) .. (271.59,442.22) ;
\draw [shift={(271.77,443.83)}, rotate = 263.34000000000003] [color={rgb, 255:red, 0; green, 0; blue, 0 }  ][line width=0.75]    (10.93,-3.29) .. controls (6.95,-1.4) and (3.31,-0.3) .. (0,0) .. controls (3.31,0.3) and (6.95,1.4) .. (10.93,3.29)   ;

\end{tikzpicture}
\caption{{The roadmap for quantum diffusion map algorithm, which also enables an efficient construction of the transition matrix for generating random walk on graphs as a byproduct. Note that we do not perform the eigen-decomposition of the transition matrix directly since quantum phase estimation (QPE) only eigen-decomposes symmetric matrices. Nevertheless the eigenpairs of the original transition matrix can be reconstructed from those of the symmetrized transition matrix.}}
\label{fig:qDM}
\end{figure}
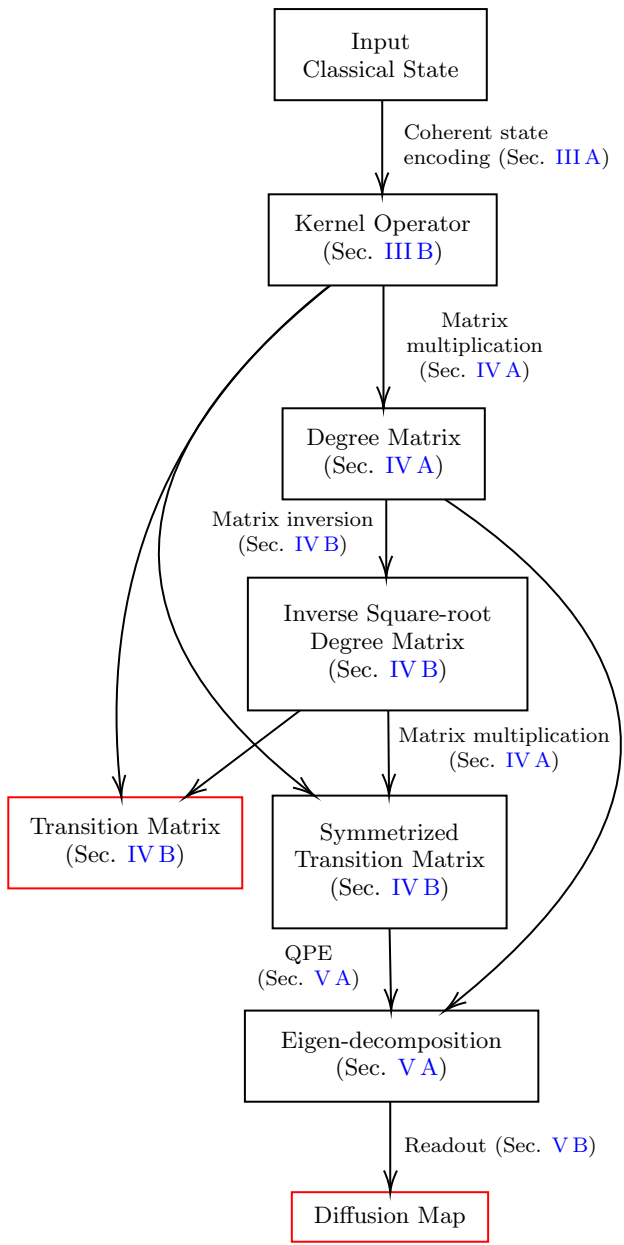

In this section, we first give an outline of qDM algorithm summarized in Fig.~\ref{fig:qDM}. Then we proceed with classical to quantum data encoding, and the construction of the kernel matrix  in Sec.~\ref{subsec: coherent} and Sec.~\ref{subsec: Kernel}, respectively. Constructing the Markov transition matrix and extracting the relevant eigen-decomposition are not straightforward, and will be discussed in the following Sec.~\ref{sec: transition_matrix} and Sec.~\ref{sec: QPE2}, respectively. 

From now on, we will focus on the case where the weighted adjacency matrix is the Gaussian kernel matrix, so that $W = K$. To reduce the dominating complexity in the construction of the diffusion map, we first turn to the idea of quantum mechanically encoding the (normalized) kernel matrix a density matrix $\hat{K}$ without explicitly evaluating each matrix element. Using the fact that the Gaussian kernel \eqref{eq: kernel} is the inner product of canonical coherent states, this can be done in $O(\log N)$ time assuming quantum random access memory (qRAM) by exploiting ``quantum parallelism", the ability to query data in superposition. {We further discuss the assumption of qRAM in Appendix~\ref{sec: complexity_kernel}.}

To construct the transition matrix, our algorithm uses as subroutines density matrix exponentiation \cite{lloyd2014quantum}, quantum matrix algebra toolbox (QMAT) \cite{zhao2019compiling}, and quantum matrix inversion \cite{harrow2009quantum,Cong2016QuantumDA}.
While it is straightforward to classically compute the degree matrix $D$ given elements of $K$, here we do not have direct access to elements of the density matrix $\hat{K}$ without performing full tomography. 
Instead, we compute $D$ via the identity
\begin{equation}\label{eq: degree_matrix}
	D  = (K {\mathbbm{1}})\odot I,
\end{equation}
where $\odot$ is element-wise multiplication also known as Hadamard product, ${\mathbbm{1}}$ is the all-ones matrix, and $I$ is the identity matrix.
Such matrix arithmetics can be done in the QMAT framework given the exponential of the relevant matrices either by density matrix exponentiation for $\hat{K}$, or Hamiltonian simulation to exponentiate sparse matrices \cite{PhysRevLett.118.010501}.
We then use quantum phase estimation (QPE) to {construct the inverse square-root of $D$ (akin to the HHL algorithm \cite{harrow2009quantum}). Then the (right) eigenvectors and eigenvalues of the transition matrix $P \equiv D^{-1}K$ can be obtained from the eigen-decomposition (via another QPE) of the symmetrized transition matrix $S\equiv D^{-1/2}KD^{-1/2}$.}
The diffusion map is then constructed classically.
The time complexity is given in Section \ref{sec:complexity_quantum}.

\subsection{Quantum state encoding into coherent states\label{subsec: coherent}}

By the fact that the Gaussian kernel arises naturally as the inner product of coherent states, ref. \cite{chatterjee2016generalized} proposed encoding classical data into multi-mode coherent states 
\begin{equation}
    {\vb*{x}}^{(i)} \mapsto \ket{\vb*{x}^{(i)}}=\bigotimes_{p=0}^{d-1}\ket{x_p^{(i)}},
    \label{eq: qData}
\end{equation}
where each single-mode coherent state $\ket{x_p^{(i)}}$ reprensents the $p^\text{th}$ feature of datum $\vb*{x}^{(i)}$.
In continuous-variable quantum systems, the canonical coherent states are states with minimal and equal uncertainties in both quadratures. They are displaced ground state of the harmonic oscillator, and can be realized, for example, as the state of classical electromagnetic radiation.
A single-mode coherent state is defined as
\begin{equation}
    \ket{x_p^{(i)}}=\exp(-\frac{\qty(x_p^{(i)})^2}{2})\sum_{n=0}^\infty \frac{\qty(x_p^{(i)})^n}{\sqrt{n!}}\ket{n},
    \label{eq: qFeature}
\end{equation}
where $\ket{n}$ is the $n^{th}$ eigenstate of the harmonic oscillator. 

The inner product between two data vectors is the Gaussian kernel
\begin{align}
\begin{split}
    \braket{\vb*{x}^{(i)}}{\vb*{x}^{(j)}} &= \prod_{p=0}^{d-1}\braket{x_p^{(i)}}{x_p^{(j)}} \\
    &= \prod_{p=0}^{d-1} \exp(-\frac{\qty(x_p^{(i)}-x_p^{(j)})^2}{2}) \\
    &= \exp(-\frac{\norm{\vb*{x}^{(i)}-\vb*{x}^{(j)}}_2^2}{2})=K\qty(\vb*{x}^{(i)},\vb*{x}^{(j)}).
\end{split}
\end{align}
Here, we take $\sigma =1$ for simplicity (though $\sigma$ can be incorporated by a scaling factor during state encoding in classical pre-processing). Recall that we focus on the kernel matrix as the weight matrix, and thus we denote $K$ for describing the weight matrix $W$.

\subsection{\label{subsec: Kernel}Kernel matrix}

Now we exploit superposition access to the encoded data to implicitly evaluate the kernel matrix. This can be done efficiently assuming qRAM for instance \cite{qRAM,QML}. Calling the oracle with the state $\frac{1}{\sqrt{N}} \sum_{i=0}^{N-1} \ket{i}$ entangles the data vectors with their labels: 
\begin{equation}
    \ket{\psi} = \frac{1}{\sqrt{N}}\sum_{i=0}^{N-1}\ket{i}_{\mathrm{label}} \otimes \ket{\vb*{x}^{(i)}}_{\mathrm{position}}.
    \label{eq: superposition}
\end{equation}
The reduced density matrix of the label space is given by the partial trace over the position space:
\begin{align}
    \Tr_\text{position}(\dyad{\psi}) &= \frac{1}{N}\sum_{i,j=0}^{N-1}\braket{\vb*{x}^{(i)}}{\vb*{x}^{(j)}}\dyad{i}{j}\nonumber\\
    &= \frac{1}{N}\sum_{i,j=0}^{N-1}K\qty(\vb*{x}^{(i)},\vb*{x}^{(j)})\dyad{i}{j}\nonumber\\
    &= \frac{K}{N} \equiv \hat{K} ,
\end{align}
where $\Tr(\hat{K})=\sum_{i=0}^{N-1}K\qty(\vb*{x}^{(i)},\vb*{x}^{(i)})=N$. Copies of $\hat{K}$ will be used to prepare the degree matrix and the transition matrix.

\section{Quantum subroutine for constructing the transition matrix\label{sec: transition_matrix}}

In this section, we show that the degree matrix and its inverse, and consequently the transition matrix, can be obtained efficiently in a coherent fashion.
In particular, if one can efficiently perform matrix multiplication and Hadamard product $\odot$ involving $\hat{K}$, then one can efficiently obtain the degree matrix according to \eqref{eq: degree_matrix}.
Such matrix arithmetics can be done in the QMAT framework \cite{zhao2019compiling} by first exponentiating the relevant (possibly non-Hermitian) matrices $A$ in the form $e^{iX_i(A)}$, where 
\begin{equation}
    X_i(A) = R_i\otimes A + R_i^{\dagger}\otimes A^{\dagger},
\end{equation}
and
\begin{align}
    R_1=\ketbra{0}{1}, && R_2=\ketbra{1}{2}, &&
    R_3=\ketbra{0}{2},
\end{align}
are qutrit operators.
Once one has the ability to apply, say, $e^{iX_1(A_1)}$ and $e^{iX_2(A_2)}$ (shuffling between $R_i,i=1,2,3$ can be done by a simple permutation of the qutrit basis states), then
using  the commutation relation between $X_1(A_1)$ and $X_2(A_2)$ one can approximate $e^{iX_3(A_3)}$ where $A_3$ is the result of the desired binary operation between $A_1$ and $A_2$. 

\subsection{\label{subsec: exponential}Exponential of degree matrix}

To compute the degree matrix, we require QMAT subroutines for (i) matrix multiplication, (ii) tensor product, and (iii) Hadamard product. The latter two subroutines reduce to matrix multiplication as $A_1 \otimes A_2 = (A_1 \otimes I) (I \otimes A_2)$, and 
\begin{equation}
    \Xi (A_1\otimes A_2) \Xi^\dagger=(A_1\odot A_2)\otimes\dyad{0},
    \label{eq: hadamard}
\end{equation}
where
\begin{align}
	\Xi = \sum_{i=0}^{N-1} \ketbra{i}{i} \otimes \ketbra{{\bf 0}}{i},
\end{align}
and $\ket{\bf 0} = \ket{0\cdots 0}$ \cite{zhao2019compiling}. 
The multiplication subroutine and its complexity is summarized in Appendix \ref{sec:QMAT}.

Define the (rank-one) density matrix
\begin{align}
    \hat{\mathbbm{1}} \equiv \frac{1}{N}\sum_{i,j=0}^{N-1}\ketbra{i}{j} = \frac{\mathbbm{1}}{N}.
\end{align}
If $e^{iX_i(\hat{K})\tau}$, $e^{iX_i(\hat{\mathbbm{1}})\tau}$, $e^{iX_i(\Xi)\tau}$ and $e^{iX_i(\Xi^{\dagger})\tau}$  can be constructed efficiently for some time $\tau$, then one can construct for a larger time $t$ (to be determined later for the purpose of phase estimation) the controlled-$e^{iX_3(D)t}$ unitary
\begin{align}
\begin{split}
    e^{iX_3(\Xi(\hat{K}\hat{\mathbbm{1}} \otimes  I)\Xi^\dagger)t} 
    &= e^{iX_3(D)t}\otimes\dyad{\vb*{0}}+I\otimes(I-\dyad{\vb*{0}})\label{eq: degree}\\
    &\equiv \mathcal{U}_D.
\end{split}
\end{align}
The last register will be discarded in the inversion step.

Now we discuss an efficient construction of the unitary matrices required in the above procedure. 
$\Xi$ is sparse and thus can be exponentiated efficiently \footnote{An $N\times N$ matrix is \emph{$s$-sparse} if there are at most $s$ nonzero entries per row. An $N\times N$ matrix is \emph{sparse} if it is at most $\mathrm{polylog}(N)$-sparse.}.
For the kernel matrix $\hat{K}$ and $\hat{\mathbbm{1}}$ which are not sparse, we can use the technique of density matrix exponentiation with time cost of  $O(\tau^2/\epsilon)$ with accuracy $\epsilon$ \cite{lloyd2014quantum}. (Note that we cannot exponentiate $X_i(A)$ directly using this method since $X_i(A)$ is not a density matrix.
{However, since $R_i + R_i^{\dagger} = \ketbra{+_i}{+_i}-\ketbra{-_i}{-_i}$, where $\ket{\pm_i}$ are the $\pm1$-eigenvectors of $R_i+R_i^\dagger$, we can separately exponentiate $\ketbra{\pm_i}{\pm_i}\otimes A$ with the time parameter $\pm\tau$ respectively and concatenate them to obtain the desired unitary operator.})
Hence, $e^{iX_3(\hat{K}\hat{\mathbbm{1}}) \tau}$ and automatically $e^{iX_3((\hat{K}\hat{\mathbbm{1}})\otimes I) \tau}$ (since $e^{iA}\otimes I = e^{iA\otimes I}$) can be created efficiently via the multiplication subroutine. 

\subsection{Inverse {square-root} of degree matrix and transition matrix}\label{subsec:invert}

Having $e^{iX_3(D)t}$ already in the exponential form, it is natural to construct the inverse {square-root} of $D$ using quantum phase estimation (QPE) akin to the HHL algorithm for matrix inversion \cite{harrow2009quantum}, with the following differences:
\begin{enumerate}
    \item Our degree matrix is always well-conditioned, having all eigenvalues in the range $1 \le d_i < N$.
    \item To construct the (symmetrized) transition matrix, we would like the resulting inverse {square-root} matrix in the form of a density matrix to be fed into the QMAT framework once more via density matrix exponentiation. Following \cite{Cong2016QuantumDA}, this implies that the function of the eigenvalues we want to compute is {$1/\sqrt[4]{d_i}$}.
\end{enumerate}

\begin{figure}
    \centering
    \includegraphics{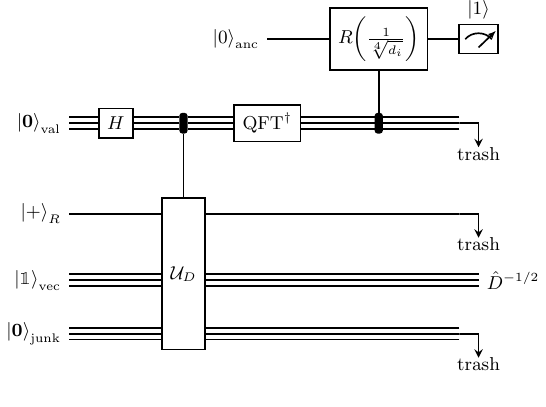}
    \caption{Quantum circuit for {square-root} degree matrix inversion. Note that QFT denotes quantum Fourier transform and $\mathcal{U}_D$ is defined by \eqref{eq: degree}.}
    \label{fig:D_inverse}
\end{figure}

The quantum circuit for the QPE (Fig.~\ref{fig:D_inverse}) is initialized in the state
\begin{equation}
    \ket{\psi_0}=\ket{\vb*{0}}_\val\otimes\ket{+}_R\otimes\ket{{\mathbbm{1}}}_\svec\otimes\ket{\vb*{0}}_\J,
    \label{eq: QPE1_initial}
\end{equation}
where $\ket{+}=\frac{1}{\sqrt{2}}(\ket{0}+\ket{2})$ and $\ket{\mathbbm{1}}=\frac{1}{\sqrt{N}}\sum_{i=0}^{N-1}\ket{i}$. The first register ``val" will eventually contain the estimated eigenvalues after the QPE, which are the components of the degree matrix. The next two registers, ``$R$" and ``vec", will contain the eigenvectors of $X_3(D)$, and the last one ``junk" is added for collapsing other terms except $e^{iX_3(D)t}$ in~(\ref{eq: degree}). 
Denote by $\kappa_D = d_\text{max}/d_\text{min}$ the condition number of $D$, i.e. the ratio of the largest eigenvalue to the smallest eigenvalue.
Setting {$t=O(\kappa_D^{0.25}/\epsilon_D)$} with accuracy $\epsilon_D$ in the QPE results in the final state $\frac{1}{\sqrt{N}}\sum_{i=0}^{N-1}\ket{\widetilde{d_i}}_\val\otimes\ket{+}_R\otimes\ket{i}_\svec\otimes\ket{\vb*{0}}_\J$, where tilde  denotes  the  binary  representation of the estimated degree. 

After the conditional rotation on the ancilla qubit ``anc", the state is 
{\begin{multline} \label{eq: conditional_rotate}
    \sum_{i=0}^{N-1}\qty(\sqrt{1-\sqrt{\frac{\check{C}}{d_i}}}\ket{0}_\text{anc}+\sqrt[4]{\frac{\check{C}}{d_i}}\ket{1}_\text{anc})\\ \otimes\ket{\widetilde{d_i}}_\val\otimes\ket{+}_R\otimes\ket{i}_\svec\otimes\ket{\vb*{0}}_\J,
\end{multline}}
\noindent where $\check{C}$ is of  order $O(d_\text{min})$. 
We then perform post-selection on the outcome ``1" 
on the ancilla qubit and take the partial trace over ``val", ``\textit{R}", and ``junk" registers to obtain
{\begin{equation}
    \hat{D}^{-1/2}=\tilde{C}\sum_{i=0}^{N-1}\sqrt{\frac{1}{d_i}}\dyad{i}{i},
\end{equation}}
where {$\tilde{C}=O\qty(\qty{\sum_{i=0}^{N-1} d_i^{-1/2}}^{-1})$} with success probability {$\Omega\qty(1/\kappa_D^{0.5})$} \cite{Cong2016QuantumDA}. 
The success probability can be amplified to near certainty by repeating {$O(\kappa_D^{0.25})$} rounds of amplitude amplification on the ancilla register \cite{harrow2009quantum}.
{Via another multiplication subroutine, we now have an access to the symmetrized transition matrix $S=D^{-1/2}KD^{-1/2}$ 
in an efficient manner.} 

{Although we will not deal with the bona fide transition matrix $P$ in the rest of this work, it is important to note that $P = D^{-1}K = D^{-1/2}D^{-1/2}K $ can be efficiently constructed via the multiplication subroutine, given that we already have an access to $D^{-1/2}$ and $K$. In particular, the exponential $e^{iX_3(P)\tau}$ can be constructed in roughly
$O(\kappa_D^{0.625} \tau^{1.5} \log^2 N)$ steps (same as $T_S(\tau)$ in Theorem \ref{thm:main}). Our algorithm thus can result in $P$ as a byproduct, which may be useful for other random-walk based algorithms.}

\section{Eigen-decomposition\label{sec: QPE2}}

{Recall that the necessary ingredients for constructing the diffusion map are
the (right) eigenvectors and the eigenvalues of the transition matrix $P = D^{-1}K$. We thus need to solve the eigenvalue problem
\begin{equation}\label{eq: nonsymmetric_eigenvec}
    D^{-1}K\ket{v_i} = \lambda_i\ket{v_i}.
\end{equation}
Quantum phase estimation (QPE) would immediately provide such an eigen-decomposition if $P$ is symmetric, which is typically not the case. Note that if one applies QPE to the hermitized $\mqty(0&P\\P^{\dagger}&0)$, one would end up with the singular vectors of $P$, rather than the desired eigenvectors of nonsymmetric $P$.
Therefore, we are in fact dealing with the generalized or nonsymmetric eigenvalue problem. 

One resolution is to transform \eqref{eq: nonsymmetric_eigenvec} to a standard symmetric eigenvalue problem
\begin{equation}\label{eq: symmetric_eigenvec}
    D^{-1/2} K D^{-1/2} \ket{s_i} = \lambda_i \ket{s_i}, 
\end{equation}
where $\ket{s_i} \equiv D^{1/2}\ket{v_i}$. 
Then, the desired eigenvectors $\ket{v_i}$ can be recovered given an efficient rotation conditional on $D^{-1/2}.$
Now QPE can be applied, the output of which will be an estimate of the eigenvalues and the eigenvectors of
\begin{equation}
    X_3(S)=(R_3+R_3^\dagger)\otimes S
\end{equation}
which are $\{\pm\lambda_i\}_{i=0}^{N-1}$ and $\ket{\pm}\otimes\ket{s_i}$. The rest of this section elaborates on the procedure for an eigen-decomposition of $S$, and on how to extract the eigenpairs of $P$ in order to construct the diffusion map classically.
}

\subsection{Transition matrix eigen-decomposition}\label{subsec: QPE3}

\begin{figure*}
    \centering
    \includegraphics{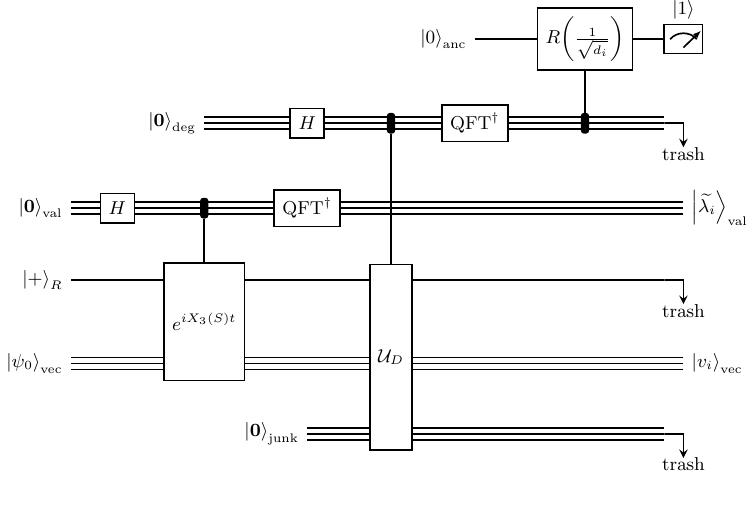}
    \caption{Quantum circuit for transition matrix $P$ eigen-decomposition. $\mathcal{U}_D$ is defined by \eqref{eq: degree}.}
    \label{fig:eigendecompose}
\end{figure*}

{The complete quantum circuit for the eigen-decomposition of $P$ is shown in Fig.~\ref{fig:eigendecompose}.
An input state is prepared in registers ``val", ``\textit{R}" and ``vec" as
\begin{equation}
    \ket{\vb*{0}}_\val\otimes\ket{+}_R\otimes\ket{\psi_0}_\svec = \ket{\vb*{0}}_\val\otimes\ket{+}_R\otimes \sum_{i=0}^{N-1}\beta_i\ket{s_i}_\svec,
    \label{eq: QPE2_initial}
\end{equation}
where $\ket{\psi_0}$ is an arbitrary initial state such that $\beta_i \equiv \braket{s_i}{\psi_0}$.
The output state of the QPE after evolving for time $t = 2\pi/\epsilon_S$, where $\epsilon_S$ is the accuracy of the estimated phases, reads
\begin{equation}
    \ket{\psi_1}_{\val,R,\svec} \equiv \sum_{i=0}^{N-1}\beta_i\ket{\widetilde{\lambda_i}}_\val\otimes\ket{+}_R\otimes\ket{s_i}_\svec,
    \label{eq: phase_estimate}
\end{equation}
where $\widetilde{\lambda_i}$ denotes the binary representation of the estimated eigenvalues in the range $[0,1)$. 

The resulting eigenvectors $\ket{s_i}$ are not yet the components of the diffusion map; they need to be conditionally rotated to the right eigenvectors $\ket{v_i} = D^{-1/2}\ket{s_i}$ of $P$ via another QPE using $e^{iX_3(D)t}$ as the time evolution operator. (Note that this step is more similar to the original HHL algorithm than our first application of QPE to construct $D^{-1/2}$. cf. \cite{Cong2016QuantumDA})
For this reason, we add to the initial state register ``deg" which will store the estimated eigenvalues $\widetilde{d_i}$ of the degree matrix, and register ``junk" to eliminate the unwanted terms in \eqref{eq: degree}:
\begin{equation}
    \ket{\vb{0}}_\text{deg}\otimes\ket{\psi_1}_{\val,R,\svec}\otimes\ket{\vb{0}}_\J.
\end{equation}
The output of the QPE after evolving for time $t = O(\kappa_D^{0.5}/\epsilon_D)$ is 
\begin{equation}
\sum_{ij=0}^{N-1}\ket{\tilde{d_j}}_\text{deg}\otimes\beta_i\ket{\widetilde{\lambda_i}}_\val\otimes\ket{+}_R\otimes(s_i)_j\ket{j}_\svec\otimes\ket{\vb{0}}_\J,
\end{equation}
where $(s_i)_j$ is the $j^\text{th}$-component of $\ket{s_i}$. After the rotation conditioning on $\widetilde{d_i}$, and $O(\kappa_D^{0.5})$ rounds of amplitude amplification on an ancilla qubit as in \eqref{eq: conditional_rotate} (not shown here), the eigenvector register is now, up to normalization, in the desired eigenstate
\begin{equation}\label{eq: QPE2_final}
    \sum_{j=0}^{N-1}\frac{(s_i)_j}{\sqrt{d_j}}\ket{j}_\svec =  D^{-1/2}\ket{s_i}_\svec = \ket{v_i}_\svec.
\end{equation}
}

\subsection{Extracting desired eigenvalues and eigenvectors\label{subsec: extract}}

Measuring the eigenvalue register ``\val" gives one of the eigenvalues and collapses the register ``\svec" to the eigenvector corresponding to that eigenvalue with probability $|\beta_i|^2$ (assuming no degeneracy as in the classical case (Sec. \ref{subsec: classical_DM})).
The eigenvector can then be extracted using a number of copies that scales almost linearly in $N$
\cite{Kerenidis_interior}. Repeating the procedure until we recover all $N-1$ distinct eigenvalues and eigenvectors allows us to classically construct the diffusion map (\ref{eq: DM}). 

{The process of recovering all eigenpairs is not deterministic; in particular, one would not be able to recover an eigenpair for which $\beta_i$ is close to zero.
What this means in practice is that one may need to change the initialization $\ket{\psi_0}$ of the QPE every so often, perhaps adaptively.
We give below the expected runtime of this process in terms of the $\beta_i$'s of a fixed fiducial state  based on the classical coupon collector's problem \cite{cCouponCollector}. Interestingly, in the best case where the $\beta_i$'s are all equal, the quantum coupon collector algorithm in \cite{qCouponCollector} guarantees that we will see all eigenpairs at least once in time $O(N)$.}

\begin{algorithm}
    \SetKwInOut{Input}{Input}\SetKwInOut{Output}{Output}
    \Input{Classical data $X$}
    \Output{Eigenvalues $\{\lambda_i\}$ and right eigenvectors $\{{v}_i\}$ of the Markov transition matrix $P$}
    \begin{enumerate}
        \item Embed $X$ into coherent state {qRAM}
        \item Perform partial trace to obtain $\hat{K}$
        \item Create state ${\hat{\mathbbm{1}}}$ and construct $e^{iX_1(\hat{K}) t}$ and $e^{iX_2({\hat{\mathbbm{1}}}) t}$
        \item Construct $e^{iX_3(D)t}$ where $t=O\qty(\kappa_D^{0.25}/\epsilon_D)$
        \item Invert {square-root} of $D$ by
        \begin{enumerate}
            \item Using $e^{iX_3(D)t}$ in QPE initialized using state (\ref{eq: QPE1_initial})
            \item Rotating the ancilla qubit controlled by the eigenvalue register ``val"
            \item Amplifying an ancilla qubit ``1" and perform partial trace on the ``val", ``\textit{R}", and ``junk" registers to get {$\hat{D}^{-1/2}$}
        \end{enumerate}
        \item Construct {$e^{iX_3(\hat{D}^{-1/2}) t}$} and $e^{iX_3(S)t}$ where $t=O(1/\epsilon_S)$
        \item Use $e^{iX_3(S)t}$ in QPE initialized using state (\ref{eq: QPE2_initial}) 
        \item {Operate $D^{-1/2}$ to obtain the eigen-decomposition of $P$ \eqref{eq: QPE2_final}}
        \item Measure the eigenvalue register ``val" to get an approximated eigenvalue $\lambda_i$ 
        \item Perform tomography to extract the eigenvector $v_i$ corresponding to the $\lambda_i$ measured in 9
    \end{enumerate}

    \caption{Quantum diffusion map 
    }\label{qDM_algo}
\end{algorithm}

\section{Time Complexity of quantum diffusion map}\label{sec:complexity_quantum}

{The complexity of qDM is summarized in Theorem \ref{thm:main} and Lemma \ref{lemma: classical_coupon}, the proof of which is provided in Appendix \ref{sec: complexity_ana}.
In Theorem \ref{thm:main}, we provide the time complexity for performing eigen-decomposition of the transition matrix as we proposed in Sec. \ref{sec: roadmap} - \ref{sec: QPE2}. In Lemma \ref{lemma: classical_coupon}, we provide the expected time complexity for reading out all eigenpairs using the classical coupon collector algorithm.}

{
\begin{theorem}\label{thm:main}

    (Time complexity of transition matrix eigen-decomposition) For a dataset $ X = \{\vb*{x}^{(i)}\}_{i=0}^{N-1}$ such that each $\vb*{x}^{(i)}\in\mathbb{R}^d$ and $d<N$, 
    steps 1-8 of Algorithm \ref{qDM_algo} constructs and eigen-decomposes the transition matrix
    with a runtime $T_E$ of
    \begin{equation}
        O\qty(T_S\qty(\frac{1}{\epsilon_S})\cdot\frac{\log N}{\epsilon_S}+T_D\qty(\frac{\kappa_D^{0.5}}{\epsilon_D})\cdot\frac{\kappa_D^{0.5}\log N}{\epsilon_D}),
        \label{eq: T_P}
    \end{equation}
    where $T_{D}$ denotes the time complexity to construct the degree matrix
    \begin{equation}
        T_D(t)=O\qty(\qty(\frac{t^{1.5}}{\epsilon_m^{0.5}}+\frac{t^{1.25}}{\epsilon_e\epsilon_m^{0.75}})\log N),
    \end{equation}    
    $T_{S}$ denotes the time complexity to construct the symmetrized transition matrix,
    \begin{widetext}
    \begin{equation}
        O\left(
        \qty(\frac{\kappa_D^{0.625}}{\epsilon_D^{2.5}\epsilon_m^{0.5}}+\frac{\kappa_D^{0.5625}}{\epsilon_D^{2.25}\epsilon_e\epsilon_m^{0.75}})
        \qty(\frac{t^{1.5}}{\epsilon_m^{0.5}\epsilon_e} +\frac{t}{\epsilon_e}) \log^2 N+\frac{t^{1.5}}{\epsilon_m^{0.5}\epsilon_e}\log N\right),
    \end{equation}
    \end{widetext}
    and $\epsilon_e$ and $\epsilon_m$ denote the errors of density matrix exponentiation and matrix multiplication, respectively.
\end{theorem}
Expressing equation~(\ref{eq: T_P}), we simplify the time complexity of the transition matrix eigen-decomposition to
\begin{equation}
    O\qty(\kappa_{D}^{0.625}\log^3{N}+\kappa_{D}^{1.25}\log^2{N}).
\end{equation}
}
{
\begin{lemma}\label{lemma: classical_coupon}
    The expected runtime to construct the diffusion map given a fixed initialization \eqref{eq: QPE2_initial} of the QPE is upper bounded by
    \begin{equation}
        \frac{\max_i\qty{|\beta_i|^2}}{\min_i\qty{|\beta_i|^2}}N \log N \cdot  O\qty(\frac{N}{\epsilon_t} \log N) \cdot T_E
    \end{equation}
    where $i\in\{0,\cdots,N-1\}$, $T_E$ denotes the time complexity of eigen-decomposing the transition matrix in Theorem \ref{thm:main} and $\epsilon_t$ denotes the error of quantum tomography.
\end{lemma}
In the special case when all the $\beta_i$'s are equal, we can recover the worst-case runtime by virtue of the recent quantum coupon collector algorithm \cite{qCouponCollector}.
\begin{lemma}\label{lemma: quantum_coupon}
    Suppose that the QPE is initialized in the state such that all the $\beta$'s are equal in \eqref{eq: QPE2_initial}. 
    Using the notations of Lemma \ref{lemma: classical_coupon}, Algorithm $\ref{qDM_algo}$ constructs the quantum diffusion map with a total runtime of
    \begin{equation}
        O\qty( \frac{N^2}{\epsilon_t}\log N \cdot T_E).
    \end{equation}
\end{lemma}
}

When the errors and the condition number are independent of the dataset size $N$, the {expected} runtime reduces to
\begin{equation}
    N^2 \text{polylog}\,N,
\end{equation}
compared to the classical diffusion map algorithm which requires $O(N^3)$ time in the worst case. In fact, the bottleneck factor of $N^2$ and the probabilistic nature of the runtime solely arises from the readout (tomography) alone (see Appendix \ref{sec: complexity_ana}), which perhaps can be circumvented if one coherently outputs the diffusion map as quantum states, providing an end-to-end computation. {Another possibility to bypass the probabilistic runtime is to extend the quantum coupon collector algorithm in Lemma \ref{lemma: quantum_coupon} to encompass a structured $\beta_i$'s and derive the worst case runtime. However, we leave this open problem for future work.} Regarding the condition number of the degree matrix, the condition number $\kappa_D$ depends on the neighborhood connectivity of the dataset, which shall be tuned to be $O(1)$ and independent of $N$, provided the scale of the neighborhood controlled by $\sigma$ in the Kernel matrix is local, such as in Fig. \ref{fig: DM_ex} (middle).

\section{Discussion and conclusion\label{sec: conclude}}

In this work, we first review classical diffusion map, an unsupervised learning algorithm for nonlinear dimensionality reduction and manifold learning.
We then explain our efforts to construct quantum algorithm for diffusion map. Our quantum diffusion map (qDM) consists of 5 major steps: coherent state data encoding scheme, a natural construction of kernel matrix from coherent states, a scheme to construct the Markov transition matrix from the kernel matrix, the eigen-decomposition of the transition matrix, and extracting relevant quantum information to construct diffusion map classically. 
The {expected} time complexity of qDM is $N^2 \text{polylog}\,N$, compared to the worst-case runtime $O(N^3)$ of a classical DM. Importantly, from accessing of qRAM to performing an eigen-decomposition of the Markov transition matrix, the total time complexity is only $O(\log^3N)$. Such exponential speedup for transition matrix construction and analysis could be useful in random walk-based algorithms.

{Note that other recently proposed quantum algorithms for manifold learning include quantum kernel principal component analysis (qKPCA) \cite{qKPCA} and quantum spectral clustering \cite{Kerenidis_PRB2021}. However, qDM has an appealing distance-preserving embedding property \eqref{eq: identity}, neither possessed by qKPCA nor quantum spectral clustering. The inner working of qDM is also starkly different from those of qKPCA and quantum spectral clustering. While qKPCA relies on approximating an arbitrary target kernel via the truncated Taylor's expansion of linear kernels \cite{qKPCA}, qDM proposes a method to {\it exactly} encode a Gaussian kernel from classical data, provided accessibility to qRAM. In addition, although both qDM and quantum spectral clustering are based on spectral graph methods, qDM manages to explicitly and efficiently construct the degree matrix inverse and the Markov transition operator, whose construction was mentioned not to be accessible efficiently \cite{Kerenidis_PRB2021}. Therefore, our work could provide a hint to achieve a quantum speedup in other algorithms that require a Markov transition operator and the analysis of its spectral properties.}

{Regarding the $N^2$ bottleneck,
a quantum algorithm for the generalized coupon collector's problem to speedup the process of finding all eigenpairs} or a method to sort the eigenvalues and select only the top ones would reduce the complexity.
Alternatively, one could devise a qDM algorithm that outputs the eigenvectors as quantum states and retain the exponential speedup. 
We leave open the question of how one could make use of such state for nonlinear dimensionality reduction.

Recently, manipulation of matrices and their spectra by block-encoding into submatrices of unitary matrices has become a topic of great interest due to its versatility in quantum algorithm design \cite{QSVT,martyn2021grand}, which may provide an alternative route to qDM algorithm.

\begin{acknowledgments}
A.S. especially thanks Dimitris Angelakis from the Centre for Quantum Technologies (CQT), Singapore, for hospitality, and for useful advice on quantum algorithms. We also thank Jirawat Tangpanitanon and Supanut Thanasilp for useful discussions. This research is supported by the Program Management Unit for Human Resources and Institutional Development, Research and Innovation (grant number B05F630108), and Sci-Super VI fund from the faculty of science, Chulalongkorn University. 
\end{acknowledgments}


\bibliography{apssamp}

\providecommand{\noopsort}[1]{}\providecommand{\singleletter}[1]{#1}%
\begin{thebibliography}{44}%
\makeatletter
\providecommand \@ifxundefined [1]{%
 \@ifx{#1\undefined}
}%
\providecommand \@ifnum [1]{%
 \ifnum #1\expandafter \@firstoftwo
 \else \expandafter \@secondoftwo
 \fi
}%
\providecommand \@ifx [1]{%
 \ifx #1\expandafter \@firstoftwo
 \else \expandafter \@secondoftwo
 \fi
}%
\providecommand \natexlab [1]{#1}%
\providecommand \enquote  [1]{``#1''}%
\providecommand \bibnamefont  [1]{#1}%
\providecommand \bibfnamefont [1]{#1}%
\providecommand \citenamefont [1]{#1}%
\providecommand \href@noop [0]{\@secondoftwo}%
\providecommand \href [0]{\begingroup \@sanitize@url \@href}%
\providecommand \@href[1]{\@@startlink{#1}\@@href}%
\providecommand \@@href[1]{\endgroup#1\@@endlink}%
\providecommand \@sanitize@url [0]{\catcode `\\12\catcode `\$12\catcode
  `\&12\catcode `\#12\catcode `\^12\catcode `\_12\catcode `\%12\relax}%
\providecommand \@@startlink[1]{}%
\providecommand \@@endlink[0]{}%
\providecommand \url  [0]{\begingroup\@sanitize@url \@url }%
\providecommand \@url [1]{\endgroup\@href {#1}{\urlprefix }}%
\providecommand \urlprefix  [0]{URL }%
\providecommand \Eprint [0]{\href }%
\providecommand \doibase [0]{https://doi.org/}%
\providecommand \selectlanguage [0]{\@gobble}%
\providecommand \bibinfo  [0]{\@secondoftwo}%
\providecommand \bibfield  [0]{\@secondoftwo}%
\providecommand \translation [1]{[#1]}%
\providecommand \BibitemOpen [0]{}%
\providecommand \bibitemStop [0]{}%
\providecommand \bibitemNoStop [0]{.\EOS\space}%
\providecommand \EOS [0]{\spacefactor3000\relax}%
\providecommand \BibitemShut  [1]{\csname bibitem#1\endcsname}%
\let\auto@bib@innerbib\@empty
\bibitem [{\citenamefont {Tenenbaum}\ \emph {et~al.}(2000)\citenamefont
  {Tenenbaum}, \citenamefont {Silva},\ and\ \citenamefont
  {Langford}}]{Tenenbaum2319}%
  \BibitemOpen
  \bibfield  {author} {\bibinfo {author} {\bibfnamefont {J.~B.}\ \bibnamefont
  {Tenenbaum}}, \bibinfo {author} {\bibfnamefont {V.~d.}\ \bibnamefont
  {Silva}},\ and\ \bibinfo {author} {\bibfnamefont {J.~C.}\ \bibnamefont
  {Langford}},\ }\bibfield  {title} {\bibinfo {title} {A global geometric
  framework for nonlinear dimensionality reduction},\ }\href
  {https://doi.org/10.1126/science.290.5500.2319} {\bibfield  {journal}
  {\bibinfo  {journal} {Science}\ }\textbf {\bibinfo {volume} {290}},\ \bibinfo
  {pages} {2319} (\bibinfo {year} {2000})}\BibitemShut {NoStop}%
\bibitem [{\citenamefont {Belkin}\ and\ \citenamefont
  {Niyogi}(2001)}]{BelkinLapEig}%
  \BibitemOpen
  \bibfield  {author} {\bibinfo {author} {\bibfnamefont {M.}~\bibnamefont
  {Belkin}}\ and\ \bibinfo {author} {\bibfnamefont {P.}~\bibnamefont
  {Niyogi}},\ }\bibfield  {title} {\bibinfo {title} {Laplacian eigenmaps and
  spectral techniques for embedding and clustering},\ }in\ \href
  {https://doi.org/10.5555/2980539.2980616} {\emph {\bibinfo {booktitle} {Adv.
  Neural Inf. Process. Syst. 14}}}\ (\bibinfo  {publisher} {MIT Press},\
  \bibinfo {year} {2001})\ pp.\ \bibinfo {pages} {585--591}\BibitemShut
  {NoStop}%
\bibitem [{\citenamefont {McInnes}\ \emph {et~al.}(2018)\citenamefont
  {McInnes}, \citenamefont {Healy}, \citenamefont {Saul},\ and\ \citenamefont
  {Großberger}}]{McInnes2018}%
  \BibitemOpen
  \bibfield  {author} {\bibinfo {author} {\bibfnamefont {L.}~\bibnamefont
  {McInnes}}, \bibinfo {author} {\bibfnamefont {J.}~\bibnamefont {Healy}},
  \bibinfo {author} {\bibfnamefont {N.}~\bibnamefont {Saul}},\ and\ \bibinfo
  {author} {\bibfnamefont {L.}~\bibnamefont {Großberger}},\ }\bibfield
  {title} {\bibinfo {title} {{UMAP}: Uniform manifold approximation and
  projection},\ }\href {https://doi.org/10.21105/joss.00861} {\bibfield
  {journal} {\bibinfo  {journal} {J. Open Source Softw.}\ }\textbf {\bibinfo
  {volume} {3}},\ \bibinfo {pages} {861} (\bibinfo {year} {2018})}\BibitemShut
  {NoStop}%
\bibitem [{\citenamefont {Scholz}\ \emph {et~al.}(2005)\citenamefont {Scholz},
  \citenamefont {Kaplan}, \citenamefont {Guy}, \citenamefont {Kopka},\ and\
  \citenamefont {Selbig}}]{Scholz_nonlinpca}%
  \BibitemOpen
  \bibfield  {author} {\bibinfo {author} {\bibfnamefont {M.}~\bibnamefont
  {Scholz}}, \bibinfo {author} {\bibfnamefont {F.}~\bibnamefont {Kaplan}},
  \bibinfo {author} {\bibfnamefont {C.~L.}\ \bibnamefont {Guy}}, \bibinfo
  {author} {\bibfnamefont {J.}~\bibnamefont {Kopka}},\ and\ \bibinfo {author}
  {\bibfnamefont {J.}~\bibnamefont {Selbig}},\ }\bibfield  {title} {\bibinfo
  {title} {{Non-linear PCA: a missing data approach}},\ }\href
  {https://doi.org/10.1093/bioinformatics/bti634} {\bibfield  {journal}
  {\bibinfo  {journal} {Bioinformatics}\ }\textbf {\bibinfo {volume} {21}},\
  \bibinfo {pages} {3887} (\bibinfo {year} {2005})}\BibitemShut {NoStop}%
\bibitem [{\citenamefont {van~der Maaten}\ and\ \citenamefont
  {Hinton}(2008)}]{tSNE_JMLR}%
  \BibitemOpen
  \bibfield  {author} {\bibinfo {author} {\bibfnamefont {L.}~\bibnamefont
  {van~der Maaten}}\ and\ \bibinfo {author} {\bibfnamefont {G.}~\bibnamefont
  {Hinton}},\ }\bibfield  {title} {\bibinfo {title} {Visualizing data using
  t-{SNE}},\ }\href {http://jmlr.org/papers/v9/vandermaaten08a.html} {\bibfield
   {journal} {\bibinfo  {journal} {J. Mach. Learn. Res.}\ }\textbf {\bibinfo
  {volume} {9}},\ \bibinfo {pages} {2579} (\bibinfo {year} {2008})}\BibitemShut
  {NoStop}%
\bibitem [{\citenamefont {Coifman}\ and\ \citenamefont
  {Lafon}(2006)}]{coifman2006diffusion}%
  \BibitemOpen
  \bibfield  {author} {\bibinfo {author} {\bibfnamefont {R.~R.}\ \bibnamefont
  {Coifman}}\ and\ \bibinfo {author} {\bibfnamefont {S.}~\bibnamefont
  {Lafon}},\ }\bibfield  {title} {\bibinfo {title} {Diffusion maps},\ }\href
  {https://doi.org/10.1016/j.acha.2006.04.006} {\bibfield  {journal} {\bibinfo
  {journal} {Appl. Comput. Harmon. Anal.}\ }\textbf {\bibinfo {volume} {21}},\
  \bibinfo {pages} {5} (\bibinfo {year} {2006})}\BibitemShut {NoStop}%
\bibitem [{\citenamefont {Lafon}(2004)}]{lafon2004diffusion}%
  \BibitemOpen
  \bibfield  {author} {\bibinfo {author} {\bibfnamefont {S.}~\bibnamefont
  {Lafon}},\ }\emph {\bibinfo {title} {Diffusion maps and geometric
  harmonics}},\ \href@noop {} {Ph.D. thesis},\ \bibinfo  {school} {Yale
  University} (\bibinfo {year} {2004})\BibitemShut {NoStop}%
\bibitem [{\citenamefont {Nadler}\ \emph {et~al.}(2006)\citenamefont {Nadler},
  \citenamefont {Lafon}, \citenamefont {Kevrekidis},\ and\ \citenamefont
  {Coifman}}]{DM_NIPS2006}%
  \BibitemOpen
  \bibfield  {author} {\bibinfo {author} {\bibfnamefont {B.}~\bibnamefont
  {Nadler}}, \bibinfo {author} {\bibfnamefont {S.}~\bibnamefont {Lafon}},
  \bibinfo {author} {\bibfnamefont {I.}~\bibnamefont {Kevrekidis}},\ and\
  \bibinfo {author} {\bibfnamefont {R.}~\bibnamefont {Coifman}},\ }\bibfield
  {title} {\bibinfo {title} {Diffusion maps, spectral clustering and
  eigenfunctions of {Fokker-Planck} operators},\ }in\ \href
  {https://doi.org/10.5555/2976248.2976368} {\emph {\bibinfo {booktitle} {Adv.
  Neural Inf. Process. Syst.}}},\ Vol.~\bibinfo {volume} {18},\ \bibinfo
  {editor} {edited by\ \bibinfo {editor} {\bibfnamefont {Y.}~\bibnamefont
  {Weiss}}, \bibinfo {editor} {\bibfnamefont {B.}~\bibnamefont
  {Sch\"{o}lkopf}},\ and\ \bibinfo {editor} {\bibfnamefont {J.}~\bibnamefont
  {Platt}}}\ (\bibinfo  {publisher} {MIT Press},\ \bibinfo {year}
  {2006})\BibitemShut {NoStop}%
\bibitem [{\citenamefont {Moon}\ \emph {et~al.}(2019)\citenamefont {Moon},
  \citenamefont {van Dijk}, \citenamefont {Wang}, \citenamefont {Gigante},
  \citenamefont {Burkhardt}, \citenamefont {Chen}, \citenamefont {Yim},
  \citenamefont {van~den Elzen}, \citenamefont {Hirn}, \citenamefont {Coifman},
  \citenamefont {Ivanova}, \citenamefont {Wolf},\ and\ \citenamefont
  {Krishnaswamy}}]{PHATE_nat2019}%
  \BibitemOpen
  \bibfield  {author} {\bibinfo {author} {\bibfnamefont {K.~R.}\ \bibnamefont
  {Moon}}, \bibinfo {author} {\bibfnamefont {D.}~\bibnamefont {van Dijk}},
  \bibinfo {author} {\bibfnamefont {Z.}~\bibnamefont {Wang}}, \bibinfo {author}
  {\bibfnamefont {S.}~\bibnamefont {Gigante}}, \bibinfo {author} {\bibfnamefont
  {D.~B.}\ \bibnamefont {Burkhardt}}, \bibinfo {author} {\bibfnamefont {W.~S.}\
  \bibnamefont {Chen}}, \bibinfo {author} {\bibfnamefont {K.}~\bibnamefont
  {Yim}}, \bibinfo {author} {\bibfnamefont {A.}~\bibnamefont {van~den Elzen}},
  \bibinfo {author} {\bibfnamefont {M.~J.}\ \bibnamefont {Hirn}}, \bibinfo
  {author} {\bibfnamefont {R.~R.}\ \bibnamefont {Coifman}}, \bibinfo {author}
  {\bibfnamefont {N.~B.}\ \bibnamefont {Ivanova}}, \bibinfo {author}
  {\bibfnamefont {G.}~\bibnamefont {Wolf}},\ and\ \bibinfo {author}
  {\bibfnamefont {S.}~\bibnamefont {Krishnaswamy}},\ }\bibfield  {title}
  {\bibinfo {title} {Visualizing structure and transitions in high-dimensional
  biological data},\ }\href {https://doi.org/10.1038/s41587-019-0336-3}
  {\bibfield  {journal} {\bibinfo  {journal} {Nat. Biotechnol.}\ }\textbf
  {\bibinfo {volume} {37}},\ \bibinfo {pages} {1482} (\bibinfo {year}
  {2019})}\BibitemShut {NoStop}%
\bibitem [{\citenamefont {Ecale~Zhou}\ \emph {et~al.}(2019)\citenamefont
  {Ecale~Zhou}, \citenamefont {Malfatti}, \citenamefont {Kimbrel},
  \citenamefont {Philipson}, \citenamefont {McNair}, \citenamefont {Hamilton},
  \citenamefont {Edwards},\ and\ \citenamefont {Souza}}]{multiPHATE_2019}%
  \BibitemOpen
  \bibfield  {author} {\bibinfo {author} {\bibfnamefont {C.~L.}\ \bibnamefont
  {Ecale~Zhou}}, \bibinfo {author} {\bibfnamefont {S.}~\bibnamefont
  {Malfatti}}, \bibinfo {author} {\bibfnamefont {J.}~\bibnamefont {Kimbrel}},
  \bibinfo {author} {\bibfnamefont {C.}~\bibnamefont {Philipson}}, \bibinfo
  {author} {\bibfnamefont {K.}~\bibnamefont {McNair}}, \bibinfo {author}
  {\bibfnamefont {T.}~\bibnamefont {Hamilton}}, \bibinfo {author}
  {\bibfnamefont {R.}~\bibnamefont {Edwards}},\ and\ \bibinfo {author}
  {\bibfnamefont {B.}~\bibnamefont {Souza}},\ }\bibfield  {title} {\bibinfo
  {title} {{multiPhATE: bioinformatics pipeline for functional annotation of
  phage isolates}},\ }\href {https://doi.org/10.1093/bioinformatics/btz258}
  {\bibfield  {journal} {\bibinfo  {journal} {Bioinformatics}\ }\textbf
  {\bibinfo {volume} {35}},\ \bibinfo {pages} {4402} (\bibinfo {year}
  {2019})}\BibitemShut {NoStop}%
\bibitem [{\citenamefont {Rodriguez-Nieva}\ and\ \citenamefont
  {Scheurer}(2019)}]{scheurer_natphys_2019}%
  \BibitemOpen
  \bibfield  {author} {\bibinfo {author} {\bibfnamefont {J.~F.}\ \bibnamefont
  {Rodriguez-Nieva}}\ and\ \bibinfo {author} {\bibfnamefont {M.~S.}\
  \bibnamefont {Scheurer}},\ }\bibfield  {title} {\bibinfo {title} {Identifying
  topological order through unsupervised machine learning},\ }\href
  {https://doi.org/10.1038/s41567-019-0512-x} {\bibfield  {journal} {\bibinfo
  {journal} {Nat. Phys.}\ }\textbf {\bibinfo {volume} {15}},\ \bibinfo {pages}
  {790} (\bibinfo {year} {2019})}\BibitemShut {NoStop}%
\bibitem [{\citenamefont {Long}\ \emph {et~al.}(2020)\citenamefont {Long},
  \citenamefont {Ren},\ and\ \citenamefont {Chen}}]{Chen_PRL_TopoPhonon2019}%
  \BibitemOpen
  \bibfield  {author} {\bibinfo {author} {\bibfnamefont {Y.}~\bibnamefont
  {Long}}, \bibinfo {author} {\bibfnamefont {J.}~\bibnamefont {Ren}},\ and\
  \bibinfo {author} {\bibfnamefont {H.}~\bibnamefont {Chen}},\ }\bibfield
  {title} {\bibinfo {title} {Unsupervised manifold clustering of topological
  phononics},\ }\href {https://doi.org/10.1103/PhysRevLett.124.185501}
  {\bibfield  {journal} {\bibinfo  {journal} {Phys. Rev. Lett.}\ }\textbf
  {\bibinfo {volume} {124}},\ \bibinfo {pages} {185501} (\bibinfo {year}
  {2020})}\BibitemShut {NoStop}%
\bibitem [{\citenamefont {Lidiak}\ and\ \citenamefont
  {Gong}(2020)}]{Gong_PRL_2020}%
  \BibitemOpen
  \bibfield  {author} {\bibinfo {author} {\bibfnamefont {A.}~\bibnamefont
  {Lidiak}}\ and\ \bibinfo {author} {\bibfnamefont {Z.}~\bibnamefont {Gong}},\
  }\bibfield  {title} {\bibinfo {title} {Unsupervised machine learning of
  quantum phase transitions using diffusion maps},\ }\href
  {https://doi.org/10.1103/PhysRevLett.125.225701} {\bibfield  {journal}
  {\bibinfo  {journal} {Phys. Rev. Lett.}\ }\textbf {\bibinfo {volume} {125}},\
  \bibinfo {pages} {225701} (\bibinfo {year} {2020})}\BibitemShut {NoStop}%
\bibitem [{\citenamefont {Kerr}\ \emph {et~al.}(2021)\citenamefont {Kerr},
  \citenamefont {Jose}, \citenamefont {Riggert},\ and\ \citenamefont
  {Mullen}}]{Kerr_PRE_2021}%
  \BibitemOpen
  \bibfield  {author} {\bibinfo {author} {\bibfnamefont {A.}~\bibnamefont
  {Kerr}}, \bibinfo {author} {\bibfnamefont {G.}~\bibnamefont {Jose}}, \bibinfo
  {author} {\bibfnamefont {C.}~\bibnamefont {Riggert}},\ and\ \bibinfo {author}
  {\bibfnamefont {K.}~\bibnamefont {Mullen}},\ }\bibfield  {title} {\bibinfo
  {title} {Automatic learning of topological phase boundaries},\ }\href
  {https://doi.org/10.1103/PhysRevE.103.023310} {\bibfield  {journal} {\bibinfo
   {journal} {Phys. Rev. E}\ }\textbf {\bibinfo {volume} {103}},\ \bibinfo
  {pages} {023310} (\bibinfo {year} {2021})}\BibitemShut {NoStop}%
\bibitem [{\citenamefont {Che}\ \emph {et~al.}(2020)\citenamefont {Che},
  \citenamefont {Gneiting}, \citenamefont {Liu},\ and\ \citenamefont
  {Nori}}]{nori_PRB_2021}%
  \BibitemOpen
  \bibfield  {author} {\bibinfo {author} {\bibfnamefont {Y.}~\bibnamefont
  {Che}}, \bibinfo {author} {\bibfnamefont {C.}~\bibnamefont {Gneiting}},
  \bibinfo {author} {\bibfnamefont {T.}~\bibnamefont {Liu}},\ and\ \bibinfo
  {author} {\bibfnamefont {F.}~\bibnamefont {Nori}},\ }\bibfield  {title}
  {\bibinfo {title} {Topological quantum phase transitions retrieved through
  unsupervised machine learning},\ }\href
  {https://doi.org/10.1103/PhysRevB.102.134213} {\bibfield  {journal} {\bibinfo
   {journal} {Phys. Rev. B}\ }\textbf {\bibinfo {volume} {102}},\ \bibinfo
  {pages} {134213} (\bibinfo {year} {2020})}\BibitemShut {NoStop}%
\bibitem [{\citenamefont {Wang}\ \emph {et~al.}(2021)\citenamefont {Wang},
  \citenamefont {Zhang}, \citenamefont {Hua},\ and\ \citenamefont
  {Wei}}]{wang_PRR_2021}%
  \BibitemOpen
  \bibfield  {author} {\bibinfo {author} {\bibfnamefont {J.}~\bibnamefont
  {Wang}}, \bibinfo {author} {\bibfnamefont {W.}~\bibnamefont {Zhang}},
  \bibinfo {author} {\bibfnamefont {T.}~\bibnamefont {Hua}},\ and\ \bibinfo
  {author} {\bibfnamefont {T.-C.}\ \bibnamefont {Wei}},\ }\bibfield  {title}
  {\bibinfo {title} {Unsupervised learning of topological phase transitions
  using the {Calinski-Harabaz} index},\ }\href
  {https://doi.org/10.1103/PhysRevResearch.3.013074} {\bibfield  {journal}
  {\bibinfo  {journal} {Phys. Rev. Research}\ }\textbf {\bibinfo {volume}
  {3}},\ \bibinfo {pages} {013074} (\bibinfo {year} {2021})}\BibitemShut
  {NoStop}%
\bibitem [{\citenamefont {Rebentrost}\ \emph {et~al.}(2018)\citenamefont
  {Rebentrost}, \citenamefont {Steffens}, \citenamefont {Marvian},\ and\
  \citenamefont {Lloyd}}]{lloyd_PRA_QSVD}%
  \BibitemOpen
  \bibfield  {author} {\bibinfo {author} {\bibfnamefont {P.}~\bibnamefont
  {Rebentrost}}, \bibinfo {author} {\bibfnamefont {A.}~\bibnamefont
  {Steffens}}, \bibinfo {author} {\bibfnamefont {I.}~\bibnamefont {Marvian}},\
  and\ \bibinfo {author} {\bibfnamefont {S.}~\bibnamefont {Lloyd}},\ }\bibfield
   {title} {\bibinfo {title} {Quantum singular-value decomposition of nonsparse
  low-rank matrices},\ }\href {https://doi.org/10.1103/PhysRevA.97.012327}
  {\bibfield  {journal} {\bibinfo  {journal} {Phys. Rev. A}\ }\textbf {\bibinfo
  {volume} {97}},\ \bibinfo {pages} {012327} (\bibinfo {year}
  {2018})}\BibitemShut {NoStop}%
\bibitem [{\citenamefont {Lloyd}\ \emph {et~al.}(2014)\citenamefont {Lloyd},
  \citenamefont {Mohseni},\ and\ \citenamefont
  {Rebentrost}}]{lloyd2014quantum}%
  \BibitemOpen
  \bibfield  {author} {\bibinfo {author} {\bibfnamefont {S.}~\bibnamefont
  {Lloyd}}, \bibinfo {author} {\bibfnamefont {M.}~\bibnamefont {Mohseni}},\
  and\ \bibinfo {author} {\bibfnamefont {P.}~\bibnamefont {Rebentrost}},\
  }\bibfield  {title} {\bibinfo {title} {Quantum principal component
  analysis},\ }\href {https://doi.org/10.1038/nphys3029} {\bibfield  {journal}
  {\bibinfo  {journal} {Nat. Phys.}\ }\textbf {\bibinfo {volume} {10}},\
  \bibinfo {pages} {631} (\bibinfo {year} {2014})}\BibitemShut {NoStop}%
\bibitem [{\citenamefont {Li}\ \emph {et~al.}(2020)\citenamefont {Li},
  \citenamefont {Zhou}, \citenamefont {Xu}, \citenamefont {Hu},\ and\
  \citenamefont {Fan}}]{qKPCA}%
  \BibitemOpen
  \bibfield  {author} {\bibinfo {author} {\bibfnamefont {Y.}~\bibnamefont
  {Li}}, \bibinfo {author} {\bibfnamefont {R.-G.}\ \bibnamefont {Zhou}},
  \bibinfo {author} {\bibfnamefont {R.}~\bibnamefont {Xu}}, \bibinfo {author}
  {\bibfnamefont {W.}~\bibnamefont {Hu}},\ and\ \bibinfo {author}
  {\bibfnamefont {P.}~\bibnamefont {Fan}},\ }\bibfield  {title} {\bibinfo
  {title} {Quantum algorithm for the nonlinear dimensionality reduction with
  arbitrary kernel},\ }\href {https://doi.org/10.1088/2058-9565/abbe66}
  {\bibfield  {journal} {\bibinfo  {journal} {Quantum Science and Technology}\
  }\textbf {\bibinfo {volume} {6}},\ \bibinfo {pages} {014001} (\bibinfo {year}
  {2020})}\BibitemShut {NoStop}%
\bibitem [{\citenamefont {Apers}\ and\ \citenamefont
  {de~Wolf}(2020)}]{dewolf_ieee_2020}%
  \BibitemOpen
  \bibfield  {author} {\bibinfo {author} {\bibfnamefont {S.}~\bibnamefont
  {Apers}}\ and\ \bibinfo {author} {\bibfnamefont {R.}~\bibnamefont
  {de~Wolf}},\ }\bibfield  {title} {\bibinfo {title} {Quantum speedup for graph
  sparsification, cut approximation and laplacian solving},\ }in\ \href
  {https://doi.org/10.1109/FOCS46700.2020.00065} {\emph {\bibinfo {booktitle}
  {Proc. IEEE 61st Annu. Symp. Found. Comput. Science (FOCS)}}}\ (\bibinfo
  {year} {2020})\ pp.\ \bibinfo {pages} {637--648}\BibitemShut {NoStop}%
\bibitem [{\citenamefont {Kerenidis}\ and\ \citenamefont
  {Landman}(2021)}]{Kerenidis_PRB2021}%
  \BibitemOpen
  \bibfield  {author} {\bibinfo {author} {\bibfnamefont {I.}~\bibnamefont
  {Kerenidis}}\ and\ \bibinfo {author} {\bibfnamefont {J.}~\bibnamefont
  {Landman}},\ }\bibfield  {title} {\bibinfo {title} {Quantum spectral
  clustering},\ }\href {https://doi.org/10.1103/PhysRevA.103.042415} {\bibfield
   {journal} {\bibinfo  {journal} {Phys. Rev. A}\ }\textbf {\bibinfo {volume}
  {103}},\ \bibinfo {pages} {042415} (\bibinfo {year} {2021})}\BibitemShut
  {NoStop}%
\bibitem [{\citenamefont {Kerenidis}\ \emph {et~al.}(2019)\citenamefont
  {Kerenidis}, \citenamefont {Landman}, \citenamefont {Luongo},\ and\
  \citenamefont {Prakash}}]{q-means_NEURIPS2019}%
  \BibitemOpen
  \bibfield  {author} {\bibinfo {author} {\bibfnamefont {I.}~\bibnamefont
  {Kerenidis}}, \bibinfo {author} {\bibfnamefont {J.}~\bibnamefont {Landman}},
  \bibinfo {author} {\bibfnamefont {A.}~\bibnamefont {Luongo}},\ and\ \bibinfo
  {author} {\bibfnamefont {A.}~\bibnamefont {Prakash}},\ }\bibfield  {title}
  {\bibinfo {title} {q-means: A quantum algorithm for unsupervised machine
  learning},\ }in\ \href
  {https://proceedings.neurips.cc/paper/2019/file/16026d60ff9b54410b3435b403afd226-Paper.pdf}
  {\emph {\bibinfo {booktitle} {Adv. Neural Inf. Process. Syst.}}},\
  Vol.~\bibinfo {volume} {32},\ \bibinfo {editor} {edited by\ \bibinfo {editor}
  {\bibfnamefont {H.}~\bibnamefont {Wallach}}, \bibinfo {editor} {\bibfnamefont
  {H.}~\bibnamefont {Larochelle}}, \bibinfo {editor} {\bibfnamefont
  {A.}~\bibnamefont {Beygelzimer}}, \bibinfo {editor} {\bibfnamefont
  {F.}~\bibnamefont {d\textquotesingle Alch\'{e}-Buc}}, \bibinfo {editor}
  {\bibfnamefont {E.}~\bibnamefont {Fox}},\ and\ \bibinfo {editor}
  {\bibfnamefont {R.}~\bibnamefont {Garnett}}}\ (\bibinfo  {publisher} {Curran
  Associates},\ \bibinfo {year} {2019})\BibitemShut {NoStop}%
\bibitem [{\citenamefont {Chatterjee}\ and\ \citenamefont
  {Yu}(2017)}]{chatterjee2016generalized}%
  \BibitemOpen
  \bibfield  {author} {\bibinfo {author} {\bibfnamefont {R.}~\bibnamefont
  {Chatterjee}}\ and\ \bibinfo {author} {\bibfnamefont {T.}~\bibnamefont
  {Yu}},\ }\bibfield  {title} {\bibinfo {title} {Generalized coherent states,
  reproducing kernels, and quantum support vector machines},\ }\href
  {https://doi.org/10.5555/3179584.3179587} {\bibfield  {journal} {\bibinfo
  {journal} {Quantum Info. Comput.}\ }\textbf {\bibinfo {volume} {17}},\
  \bibinfo {pages} {1292–1306} (\bibinfo {year} {2017})}\BibitemShut
  {NoStop}%
\bibitem [{\citenamefont {Dua}\ and\ \citenamefont {Graff}(2017)}]{Dua:2019}%
  \BibitemOpen
  \bibfield  {author} {\bibinfo {author} {\bibfnamefont {D.}~\bibnamefont
  {Dua}}\ and\ \bibinfo {author} {\bibfnamefont {C.}~\bibnamefont {Graff}},\
  }\href {http://archive.ics.uci.edu/ml} {\bibinfo {title} {{UCI} machine
  learning repository}} (\bibinfo {year} {2017})\BibitemShut {NoStop}%
\bibitem [{\citenamefont {Trefethen}\ and\ \citenamefont
  {Bau~III}(1997)}]{trefethen1997numerical}%
  \BibitemOpen
  \bibfield  {author} {\bibinfo {author} {\bibfnamefont {L.~N.}\ \bibnamefont
  {Trefethen}}\ and\ \bibinfo {author} {\bibfnamefont {D.}~\bibnamefont
  {Bau~III}},\ }\href@noop {} {\emph {\bibinfo {title} {Numerical linear
  algebra}}},\ Vol.~\bibinfo {volume} {50}\ (\bibinfo  {publisher} {Siam},\
  \bibinfo {year} {1997})\BibitemShut {NoStop}%
\bibitem [{\citenamefont {Zhao}\ \emph {et~al.}(2021)\citenamefont {Zhao},
  \citenamefont {Zhao}, \citenamefont {Rebentrost},\ and\ \citenamefont
  {Fitzsimons}}]{zhao2019compiling}%
  \BibitemOpen
  \bibfield  {author} {\bibinfo {author} {\bibfnamefont {L.}~\bibnamefont
  {Zhao}}, \bibinfo {author} {\bibfnamefont {Z.}~\bibnamefont {Zhao}}, \bibinfo
  {author} {\bibfnamefont {P.}~\bibnamefont {Rebentrost}},\ and\ \bibinfo
  {author} {\bibfnamefont {J.}~\bibnamefont {Fitzsimons}},\ }\bibfield  {title}
  {\bibinfo {title} {Compiling basic linear algebra subroutines for quantum
  computers},\ }\href {https://doi.org/10.1007/s42484-021-00048-8} {\bibfield
  {journal} {\bibinfo  {journal} {Quantum Mach. Intell.}\ }\textbf {\bibinfo
  {volume} {3}},\ \bibinfo {pages} {21} (\bibinfo {year} {2021})}\BibitemShut
  {NoStop}%
\bibitem [{\citenamefont {Harrow}\ \emph {et~al.}(2009)\citenamefont {Harrow},
  \citenamefont {Hassidim},\ and\ \citenamefont {Lloyd}}]{harrow2009quantum}%
  \BibitemOpen
  \bibfield  {author} {\bibinfo {author} {\bibfnamefont {A.~W.}\ \bibnamefont
  {Harrow}}, \bibinfo {author} {\bibfnamefont {A.}~\bibnamefont {Hassidim}},\
  and\ \bibinfo {author} {\bibfnamefont {S.}~\bibnamefont {Lloyd}},\ }\bibfield
   {title} {\bibinfo {title} {Quantum algorithm for linear systems of
  equations},\ }\href {https://doi.org/10.1103/PhysRevLett.103.150502}
  {\bibfield  {journal} {\bibinfo  {journal} {Phys. Rev. Lett.}\ }\textbf
  {\bibinfo {volume} {103}},\ \bibinfo {pages} {150502} (\bibinfo {year}
  {2009})}\BibitemShut {NoStop}%
\bibitem [{\citenamefont {Cong}\ and\ \citenamefont
  {Duan}(2016)}]{Cong2016QuantumDA}%
  \BibitemOpen
  \bibfield  {author} {\bibinfo {author} {\bibfnamefont {I.}~\bibnamefont
  {Cong}}\ and\ \bibinfo {author} {\bibfnamefont {L.}~\bibnamefont {Duan}},\
  }\bibfield  {title} {\bibinfo {title} {Quantum discriminant analysis for
  dimensionality reduction and classification},\ }\href
  {https://doi.org/10.1088/1367-2630/18/7/073011} {\bibfield  {journal}
  {\bibinfo  {journal} {New J. Phys.}\ }\textbf {\bibinfo {volume} {18}},\
  \bibinfo {pages} {073011} (\bibinfo {year} {2016})}\BibitemShut {NoStop}%
\bibitem [{\citenamefont {Low}\ and\ \citenamefont
  {Chuang}(2017)}]{PhysRevLett.118.010501}%
  \BibitemOpen
  \bibfield  {author} {\bibinfo {author} {\bibfnamefont {G.~H.}\ \bibnamefont
  {Low}}\ and\ \bibinfo {author} {\bibfnamefont {I.~L.}\ \bibnamefont
  {Chuang}},\ }\bibfield  {title} {\bibinfo {title} {Optimal hamiltonian
  simulation by quantum signal processing},\ }\href
  {https://doi.org/10.1103/PhysRevLett.118.010501} {\bibfield  {journal}
  {\bibinfo  {journal} {Phys. Rev. Lett.}\ }\textbf {\bibinfo {volume} {118}},\
  \bibinfo {pages} {010501} (\bibinfo {year} {2017})}\BibitemShut {NoStop}%
\bibitem [{\citenamefont {Giovannetti}\ \emph {et~al.}(2008)\citenamefont
  {Giovannetti}, \citenamefont {Lloyd},\ and\ \citenamefont {Maccone}}]{qRAM}%
  \BibitemOpen
  \bibfield  {author} {\bibinfo {author} {\bibfnamefont {V.}~\bibnamefont
  {Giovannetti}}, \bibinfo {author} {\bibfnamefont {S.}~\bibnamefont {Lloyd}},\
  and\ \bibinfo {author} {\bibfnamefont {L.}~\bibnamefont {Maccone}},\
  }\bibfield  {title} {\bibinfo {title} {Quantum random access memory},\ }\href
  {https://doi.org/10.1103/PhysRevLett.100.160501} {\bibfield  {journal}
  {\bibinfo  {journal} {Phys. Rev. Lett.}\ }\textbf {\bibinfo {volume} {100}},\
  \bibinfo {pages} {160501} (\bibinfo {year} {2008})}\BibitemShut {NoStop}%
\bibitem [{\citenamefont {Biamonte}\ \emph {et~al.}(2017)\citenamefont
  {Biamonte}, \citenamefont {Wittek}, \citenamefont {Pancotti}, \citenamefont
  {Rebentrost}, \citenamefont {Wiebe},\ and\ \citenamefont {Lloyd}}]{QML}%
  \BibitemOpen
  \bibfield  {author} {\bibinfo {author} {\bibfnamefont {J.}~\bibnamefont
  {Biamonte}}, \bibinfo {author} {\bibfnamefont {P.}~\bibnamefont {Wittek}},
  \bibinfo {author} {\bibfnamefont {N.}~\bibnamefont {Pancotti}}, \bibinfo
  {author} {\bibfnamefont {P.}~\bibnamefont {Rebentrost}}, \bibinfo {author}
  {\bibfnamefont {N.}~\bibnamefont {Wiebe}},\ and\ \bibinfo {author}
  {\bibfnamefont {S.}~\bibnamefont {Lloyd}},\ }\bibfield  {title} {\bibinfo
  {title} {Quantum machine learning},\ }\href
  {https://doi.org/10.1038/nature23474} {\bibfield  {journal} {\bibinfo
  {journal} {Nature}\ }\textbf {\bibinfo {volume} {549}},\ \bibinfo {pages}
  {195} (\bibinfo {year} {2017})}\BibitemShut {NoStop}%
\bibitem [{Note1()}]{Note1}%
  \BibitemOpen
  \bibinfo {note} {An $N\times N$ matrix is \protect \emph {$s$-sparse} if
  there are at most $s$ nonzero entries per row. An $N\times N$ matrix is
  \protect \emph {sparse} if it is at most $\protect \mathrm
  {polylog}(N)$-sparse.}\BibitemShut {Stop}%
\bibitem [{\citenamefont {Kerenidis}\ and\ \citenamefont
  {Prakash}(2020)}]{Kerenidis_interior}%
  \BibitemOpen
  \bibfield  {author} {\bibinfo {author} {\bibfnamefont {I.}~\bibnamefont
  {Kerenidis}}\ and\ \bibinfo {author} {\bibfnamefont {A.}~\bibnamefont
  {Prakash}},\ }\bibfield  {title} {\bibinfo {title} {A quantum interior point
  method for lps and sdps},\ }\href {https://doi.org/10.1145/3406306}
  {\bibfield  {journal} {\bibinfo  {journal} {ACM Trans. Quantum Comput.}\
  }\textbf {\bibinfo {volume} {1}},\ \bibinfo {pages} {1} (\bibinfo {year}
  {2020})}\BibitemShut {NoStop}%
\bibitem [{\citenamefont {Shank}\ and\ \citenamefont
  {Yang}(2013)}]{cCouponCollector}%
  \BibitemOpen
  \bibfield  {author} {\bibinfo {author} {\bibfnamefont {N.~B.}\ \bibnamefont
  {Shank}}\ and\ \bibinfo {author} {\bibfnamefont {H.}~\bibnamefont {Yang}},\
  }\bibfield  {title} {\bibinfo {title} {Coupon collector problem for
  non-uniform coupons and random quotas},\ }\href
  {https://doi.org/10.37236/3348} {\bibfield  {journal} {\bibinfo  {journal}
  {The Electronic Journal of Combinatorics}\ }\textbf {\bibinfo {volume}
  {20}},\ \bibinfo {pages} {P33} (\bibinfo {year} {2013})}\BibitemShut
  {NoStop}%
\bibitem [{\citenamefont {Arunachalam}\ \emph {et~al.}(2020)\citenamefont
  {Arunachalam}, \citenamefont {Belovs}, \citenamefont {Childs}, \citenamefont
  {Kothari}, \citenamefont {Rosmanis},\ and\ \citenamefont
  {de~Wolf}}]{qCouponCollector}%
  \BibitemOpen
  \bibfield  {author} {\bibinfo {author} {\bibfnamefont {S.}~\bibnamefont
  {Arunachalam}}, \bibinfo {author} {\bibfnamefont {A.}~\bibnamefont {Belovs}},
  \bibinfo {author} {\bibfnamefont {A.~M.}\ \bibnamefont {Childs}}, \bibinfo
  {author} {\bibfnamefont {R.}~\bibnamefont {Kothari}}, \bibinfo {author}
  {\bibfnamefont {A.}~\bibnamefont {Rosmanis}},\ and\ \bibinfo {author}
  {\bibfnamefont {R.}~\bibnamefont {de~Wolf}},\ }\bibfield  {title} {\bibinfo
  {title} {{Quantum Coupon Collector}},\ }in\ \href
  {https://doi.org/10.4230/LIPIcs.TQC.2020.10} {\emph {\bibinfo {booktitle}
  {15th Conference on the Theory of Quantum Computation, Communication and
  Cryptography (TQC 2020)}}},\ \bibinfo {series} {Leibniz International
  Proceedings in Informatics (LIPIcs)}, Vol.\ \bibinfo {volume} {158},\
  \bibinfo {editor} {edited by\ \bibinfo {editor} {\bibfnamefont {S.~T.}\
  \bibnamefont {Flammia}}}\ (\bibinfo  {publisher} {Schloss
  Dagstuhl--Leibniz-Zentrum f{\"u}r Informatik},\ \bibinfo {address} {Dagstuhl,
  Germany},\ \bibinfo {year} {2020})\ pp.\ \bibinfo {pages}
  {10:1--10:17}\BibitemShut {NoStop}%
\bibitem [{\citenamefont {Gily\'{e}n}\ \emph {et~al.}(2019)\citenamefont
  {Gily\'{e}n}, \citenamefont {Su}, \citenamefont {Low},\ and\ \citenamefont
  {Wiebe}}]{QSVT}%
  \BibitemOpen
  \bibfield  {author} {\bibinfo {author} {\bibfnamefont {A.}~\bibnamefont
  {Gily\'{e}n}}, \bibinfo {author} {\bibfnamefont {Y.}~\bibnamefont {Su}},
  \bibinfo {author} {\bibfnamefont {G.~H.}\ \bibnamefont {Low}},\ and\ \bibinfo
  {author} {\bibfnamefont {N.}~\bibnamefont {Wiebe}},\ }\bibfield  {title}
  {\bibinfo {title} {Quantum singular value transformation and beyond:
  Exponential improvements for quantum matrix arithmetics},\ }in\ \href
  {https://doi.org/10.1145/3313276.3316366} {\emph {\bibinfo {booktitle} {Proc.
  51st Annu. ACM SIGACT Symp. Theor. Comput.}}},\ \bibinfo {series and number}
  {STOC 2019}\ (\bibinfo  {publisher} {Association for Computing Machinery},\
  \bibinfo {address} {New York, NY, USA},\ \bibinfo {year} {2019})\ p.\
  \bibinfo {pages} {193–204}\BibitemShut {NoStop}%
\bibitem [{\citenamefont {Martyn}\ \emph {et~al.}(2021)\citenamefont {Martyn},
  \citenamefont {Rossi}, \citenamefont {Tan},\ and\ \citenamefont
  {Chuang}}]{martyn2021grand}%
  \BibitemOpen
  \bibfield  {author} {\bibinfo {author} {\bibfnamefont {J.~M.}\ \bibnamefont
  {Martyn}}, \bibinfo {author} {\bibfnamefont {Z.~M.}\ \bibnamefont {Rossi}},
  \bibinfo {author} {\bibfnamefont {A.~K.}\ \bibnamefont {Tan}},\ and\ \bibinfo
  {author} {\bibfnamefont {I.~L.}\ \bibnamefont {Chuang}},\ }\bibfield  {title}
  {\bibinfo {title} {A grand unification of quantum algorithms},\ }\href
  {https://arxiv.org/abs/2105.02859} {\bibfield  {journal} {\bibinfo  {journal}
  {arXiv:2105.02859}\ } (\bibinfo {year} {2021})}\BibitemShut {NoStop}%
\bibitem [{\citenamefont {Rebentrost}\ \emph {et~al.}(2014)\citenamefont
  {Rebentrost}, \citenamefont {Mohseni},\ and\ \citenamefont
  {Lloyd}}]{PhysRevLett.113.130503}%
  \BibitemOpen
  \bibfield  {author} {\bibinfo {author} {\bibfnamefont {P.}~\bibnamefont
  {Rebentrost}}, \bibinfo {author} {\bibfnamefont {M.}~\bibnamefont
  {Mohseni}},\ and\ \bibinfo {author} {\bibfnamefont {S.}~\bibnamefont
  {Lloyd}},\ }\bibfield  {title} {\bibinfo {title} {Quantum support vector
  machine for big data classification},\ }\href
  {https://doi.org/10.1103/PhysRevLett.113.130503} {\bibfield  {journal}
  {\bibinfo  {journal} {Phys. Rev. Lett.}\ }\textbf {\bibinfo {volume} {113}},\
  \bibinfo {pages} {130503} (\bibinfo {year} {2014})}\BibitemShut {NoStop}%
\bibitem [{\citenamefont {Yamasaki}\ \emph {et~al.}(2020)\citenamefont
  {Yamasaki}, \citenamefont {Subramanian}, \citenamefont {Sonoda},\ and\
  \citenamefont {Koashi}}]{yamasaki2020learning}%
  \BibitemOpen
  \bibfield  {author} {\bibinfo {author} {\bibfnamefont {H.}~\bibnamefont
  {Yamasaki}}, \bibinfo {author} {\bibfnamefont {S.}~\bibnamefont
  {Subramanian}}, \bibinfo {author} {\bibfnamefont {S.}~\bibnamefont
  {Sonoda}},\ and\ \bibinfo {author} {\bibfnamefont {M.}~\bibnamefont
  {Koashi}},\ }\bibfield  {title} {\bibinfo {title} {Learning with optimized
  random features: Exponential speedup by quantum machine learning without
  sparsity and low-rank assumptions},\ }\href
  {https://arxiv.org/abs/2004.10756} {\bibfield  {journal} {\bibinfo  {journal}
  {arXiv:2004.10756}\ } (\bibinfo {year} {2020})}\BibitemShut {NoStop}%
\bibitem [{\citenamefont {Jiang}\ \emph {et~al.}(2019)\citenamefont {Jiang},
  \citenamefont {Pu}, \citenamefont {Chang}, \citenamefont {Li}, \citenamefont
  {Zhang},\ and\ \citenamefont {Duan}}]{qRAM_expe1}%
  \BibitemOpen
  \bibfield  {author} {\bibinfo {author} {\bibfnamefont {N.}~\bibnamefont
  {Jiang}}, \bibinfo {author} {\bibfnamefont {Y.-F.}\ \bibnamefont {Pu}},
  \bibinfo {author} {\bibfnamefont {W.}~\bibnamefont {Chang}}, \bibinfo
  {author} {\bibfnamefont {C.}~\bibnamefont {Li}}, \bibinfo {author}
  {\bibfnamefont {S.}~\bibnamefont {Zhang}},\ and\ \bibinfo {author}
  {\bibfnamefont {L.-M.}\ \bibnamefont {Duan}},\ }\bibfield  {title} {\bibinfo
  {title} {Experimental realization of 105-qubit random access quantum
  memory},\ }\href {https://doi.org/10.1038/s41534-019-0144-0} {\bibfield
  {journal} {\bibinfo  {journal} {npj Quantum Inf.}\ }\textbf {\bibinfo
  {volume} {5}},\ \bibinfo {pages} {1} (\bibinfo {year} {2019})}\BibitemShut
  {NoStop}%
\bibitem [{\citenamefont {Hann}\ \emph {et~al.}(2019)\citenamefont {Hann},
  \citenamefont {Zou}, \citenamefont {Zhang}, \citenamefont {Chu},
  \citenamefont {Schoelkopf}, \citenamefont {Girvin},\ and\ \citenamefont
  {Jiang}}]{qRAM_expe2}%
  \BibitemOpen
  \bibfield  {author} {\bibinfo {author} {\bibfnamefont {C.~T.}\ \bibnamefont
  {Hann}}, \bibinfo {author} {\bibfnamefont {C.-L.}\ \bibnamefont {Zou}},
  \bibinfo {author} {\bibfnamefont {Y.}~\bibnamefont {Zhang}}, \bibinfo
  {author} {\bibfnamefont {Y.}~\bibnamefont {Chu}}, \bibinfo {author}
  {\bibfnamefont {R.~J.}\ \bibnamefont {Schoelkopf}}, \bibinfo {author}
  {\bibfnamefont {S.~M.}\ \bibnamefont {Girvin}},\ and\ \bibinfo {author}
  {\bibfnamefont {L.}~\bibnamefont {Jiang}},\ }\bibfield  {title} {\bibinfo
  {title} {Hardware-efficient quantum random access memory with hybrid quantum
  acoustic systems},\ }\href {https://doi.org/10.1103/PhysRevLett.123.250501}
  {\bibfield  {journal} {\bibinfo  {journal} {Phys. Rev. Lett.}\ }\textbf
  {\bibinfo {volume} {123}},\ \bibinfo {pages} {250501} (\bibinfo {year}
  {2019})}\BibitemShut {NoStop}%
\bibitem [{\citenamefont {Di~Matteo}\ \emph {et~al.}(2020)\citenamefont
  {Di~Matteo}, \citenamefont {Gheorghiu},\ and\ \citenamefont
  {Mosca}}]{qRAM_estimate}%
  \BibitemOpen
  \bibfield  {author} {\bibinfo {author} {\bibfnamefont {O.}~\bibnamefont
  {Di~Matteo}}, \bibinfo {author} {\bibfnamefont {V.}~\bibnamefont
  {Gheorghiu}},\ and\ \bibinfo {author} {\bibfnamefont {M.}~\bibnamefont
  {Mosca}},\ }\bibfield  {title} {\bibinfo {title} {Fault-tolerant resource
  estimation of quantum random-access memories},\ }\href
  {https://doi.org/10.1109/TQE.2020.2965803} {\bibfield  {journal} {\bibinfo
  {journal} {IEEE Trans. Quantum Eng.}\ }\textbf {\bibinfo {volume} {1}},\
  \bibinfo {pages} {1} (\bibinfo {year} {2020})}\BibitemShut {NoStop}%
\bibitem [{\citenamefont {Hann}\ \emph {et~al.}(2021)\citenamefont {Hann},
  \citenamefont {Lee}, \citenamefont {Girvin},\ and\ \citenamefont
  {Jiang}}]{qRAM_resilience}%
  \BibitemOpen
  \bibfield  {author} {\bibinfo {author} {\bibfnamefont {C.~T.}\ \bibnamefont
  {Hann}}, \bibinfo {author} {\bibfnamefont {G.}~\bibnamefont {Lee}}, \bibinfo
  {author} {\bibfnamefont {S.}~\bibnamefont {Girvin}},\ and\ \bibinfo {author}
  {\bibfnamefont {L.}~\bibnamefont {Jiang}},\ }\bibfield  {title} {\bibinfo
  {title} {Resilience of quantum random access memory to generic noise},\
  }\href {https://doi.org/10.1103/PRXQuantum.2.020311} {\bibfield  {journal}
  {\bibinfo  {journal} {PRX Quantum}\ }\textbf {\bibinfo {volume} {2}},\
  \bibinfo {pages} {020311} (\bibinfo {year} {2021})}\BibitemShut {NoStop}%
\bibitem [{\citenamefont {Dervovic}\ \emph {et~al.}(2018)\citenamefont
  {Dervovic}, \citenamefont {Herbster}, \citenamefont {Mountney}, \citenamefont
  {Severini}, \citenamefont {Usher},\ and\ \citenamefont
  {Wossnig}}]{dervovic2018HHLprimer}%
  \BibitemOpen
  \bibfield  {author} {\bibinfo {author} {\bibfnamefont {D.}~\bibnamefont
  {Dervovic}}, \bibinfo {author} {\bibfnamefont {M.}~\bibnamefont {Herbster}},
  \bibinfo {author} {\bibfnamefont {P.}~\bibnamefont {Mountney}}, \bibinfo
  {author} {\bibfnamefont {S.}~\bibnamefont {Severini}}, \bibinfo {author}
  {\bibfnamefont {N.}~\bibnamefont {Usher}},\ and\ \bibinfo {author}
  {\bibfnamefont {L.}~\bibnamefont {Wossnig}},\ }\bibfield  {title} {\bibinfo
  {title} {Quantum linear systems algorithms: a primer},\ }\href
  {https://arxiv.org/abs/1802.08227} {\bibfield  {journal} {\bibinfo  {journal}
  {arXiv:1802.08227}\ } (\bibinfo {year} {2018})}\BibitemShut {NoStop}%
\end{thebibliography}%

\appendix

\section{QMAT matrix multiplication subroutine \label{sec:QMAT}}

As explained in Sec. \ref{subsec: exponential}, all QMAT subroutines \cite{zhao2019compiling} utilized in our work reduce to that of matrix multiplication, which we now describe. Given matrices $A_1$ and $A_2$, the idea of the multiplication subroutine is to construct a Hermitian embedding of the product $A_1 A_2$,
\begin{equation*}
    X_3(A_1 A_2) =  \mqty(0&0&A_1A_2\\0&0&0\\(A_1A_2)^{\dagger}&0&0),
\end{equation*}
from the following commutator:
\begin{equation}
    X_3(A_1 A_2) = U X_3(iA_1 A_2) U^{\dagger} 
    = i U [X_1(A_1), X_2(A_2)] U^{\dagger},
\end{equation}
where
\begin{equation*}
    U=\mqty(\sqrt{-i}I&0&0\\0&I&0\\0&0&\sqrt{i}I).
\end{equation*}
Thus, the goal is to simulate the commutator of two hermitian operators. This can be accomplished with bounded error via the second order of the Baker-Campbell-Hausdorff formula. Define
\begin{equation*}
    \tilde{l}(x,y)\equiv e^x e^y e^{-x}e^{-y} e^{-x}e^{-y}e^x e^y.
\end{equation*}
With the step size $t/m$, we have that \cite{zhao2019compiling},
\begin{equation}
    \tilde{l}(xt/m, yt/m) = e^{2[x,y]t^2/m^2 + O(t^4/m^4)}.
\end{equation}
Thus, iterating the above $n' = m^2/(2t) = O(t^2/\epsilon_m)$ times, 
\begin{equation}
    \tilde{l}\left( x t/m, y t/m \right)^{n^2/(2t)} = e^{[x,y]t + O(t^3/m^2)},
\end{equation}
correctly simulates $e^{[x,y]t}$ up to an error $\epsilon_m = O(t^3/m^2)$.

Putting all these together, the QMAT multiplication subroutine consumes $n' = m^2/(2t)$ copies of $e^{iX_1(A_1)t/m}$ and $e^{iX_2(A_2)t/m}$ (and their inverses), and outputs a unitary operator
\begin{equation}
    U[l(iX_1(A_1)t/m,iX_2(A_2)t/m)]^{n^{\prime}} U^\dagger,
    \label{eq:exp_multi}
\end{equation}
that approximates $e^{iX_3(A_1A_2)}$ up to an error $\epsilon_m$ in the operator norm.

\section{Complexity analysis\label{sec: complexity_ana}}
This section describes the calculation for the time complexity that leads to the result in Theorem \ref{thm:main}. We separate the algorithm into sequential steps as we have done with the classical DM (Sec.~\ref{subsec: classical complexity}), namely, the embedding of classical data into the kernel $\hat{K}$, the calculation of the transition matrix $P$, and the calculation of the eigen-decomposition.

\subsection{Preparing the kernel}\label{sec: complexity_kernel}
In creating $\hat{K}$, we consider only the time complexity coming from storing classical data into {qRAM} (Sec. \ref{subsec: coherent}) as partial trace does not create additional time cost.
{Here we have assumed a common position among researchers in the field that encoding of classical data can be made efficient via the existence of an oracle \cite{lloyd2014quantum, PhysRevLett.113.130503, Cong2016QuantumDA, q-means_NEURIPS2019,yamasaki2020learning,QML}, and analyze the algorithm based on this assumption.
Note that practical feasibility and resource estimate of implementing qRAM is still an ongoing research topic~\cite{QML,qRAM_expe1,qRAM_expe2,qRAM_estimate,qRAM_resilience}.}
Assuming that the data consists of $N$ vectors each with $d$ features, the data can be encoded into a {qRAM} of the size $dN$, which can be accessed using $O(\log(dN))$ operations (\ref{eq: superposition}) \cite{PhysRevLett.113.130503}.
We assume here that we are dealing with big data, i.e., $d \le N$. Therefore, the time cost becomes $O(\log N^2)=O(\log N)$.

\subsection{Calculating the transition matrix}\label{sec: complexity_S}

Next, we construct the exponential of $P$.
We start with exponentiating the necessary matrices.
Then we multiply these matrices in order to create $\text{e}^{\imath X_3(D)t}$. Next, we inverse the degree matrix $D$ before constructing the transition matrix $P=D^{-1}K$.

The time cost of creating exponentials depends on whether the quantum simulation of matrices or quantum state is used.
Creating exponentials from multiple copies of density matrices, i.e., $e^{iX_3(\hat{K})t}$ and $e^{iX_3(\hat{{\mathbbm{1}}})t}$ , results in a time complexity of $O(t^2/\epsilon_e)$, where $\epsilon_e$ is the accuracy for the exponentiation~\cite{lloyd2014quantum}.
Combined with the number of operations for storing the states, which is $O(\log N)$, the time complexity is $O(t^2/\epsilon_e \log N)$.

Exponentials of sparse matrices can be simulated using the method in \cite{PhysRevLett.118.010501} with query complexity of $O\qty(t+\frac{\log(1/\epsilon_s)}{\log\log(1/\epsilon_s)})$, where $\epsilon_s$ is the accuracy of the method. For the matrices $e^{iX_3(\Xi)t}$ and $e^{iX_3(\Xi^\dagger)t}$, we also need to take into account the gate complexity $O(\log N)$. Thus the time complexity for preparing each of these exponentials is $O(t\log N)$. 

To construct the degree matrix, we use the matrix multiplication subroutine in Appendix \ref{sec:QMAT}. This subroutine uses 4 copies of each input exponential, which is multiplied and then applied $n^\prime$ times.
In creating $e^{iX_3(\hat{K}{\mathbbm{1}}\otimes I)}$ (Sec. \ref{subsec: exponential}), the time cost comes from multiplying the exponential of $\hat{K}$ and of ${\mathbbm{1}}$ whereas $\otimes I$ is assumed to be a free operation.
Therefore, the time cost is $O\qty(n^\prime\qty(\frac{t^2}{m^2\epsilon_e}\log N))$. Given the condition that $n^\prime=\frac{m^{2}}{2t}=O\left(\frac{t^2}{\epsilon_m}\right)~$\cite{zhao2019compiling}, we rewrite the time cost to $O(t\log N\epsilon_e^{-1})$. 

In the next two steps, we multiply $\Xi$ and $\Xi^\dagger$ into $K{\mathbbm{1}}\otimes I$. Again, we apply the multiplication subroutine iteratively with 4 copies of the result from the previous multiplication as one of the input.
These steps admit additional order of $\epsilon_m$ and $t$ which leads the time cost of $e^{iX_3(D)t}$ to be
\begin{equation}
\label{eq:T_degree}
    T_D(t)=O\qty(\qty(\frac{t^{1.5}}{\epsilon_m^{0.5}}+\frac{t^{1.25}}{\epsilon_e\epsilon_m^{0.75}})\log N).
\end{equation}

Next, the degree matrix $D$ is inverted (Sec. \ref{subsec:invert}) by first performing QPE that takes the prepared $e^{iX_3(D)t}$ as the unitary operator in QPE then a controlled rotation. The time cost of the QPE is $O(T_D/\epsilon_D\log m)$~\cite{dervovic2018HHLprimer}, where $m$ is the number of eigenvalues which is $m=N$, and $\epsilon_D$ is the error of estimated phase.
Assuming the controlled rotation is executed in time $O(1)$, we have accumulated a time cost of
$T_D+O(T_D/\epsilon_D\log N)+O(1)$. The time complexity is thus dominated by $O(T_D/\epsilon_D\log N)$.

The time $T_D$~(\ref{eq:T_degree}) is a function of $t$, which we now find through error analysis of matrix inversion.
As we are inverting the eigenvalues of $D$ to {a function of $1/\sqrt[4]{x}$} \cite{Cong2016QuantumDA}, the error after conditional rotation becomes {$\epsilon_D=O(1/\sqrt[4]{d_i}t)<O(\kappa_D^{0.25}/t)$. Thus $t=O(\kappa_D^{0.25}/\epsilon_D)$. }

As the success probability of measuring $\ket{1}$ in the ancilla qubit is {$\Omega(1/\kappa_D^{0.5})$, the number of rounds necessary to obtain a successful result is $O(\kappa_D^{0.5})$.
To accelerate the process, we use the amplitude amplification to amplify the success probability in $O(\kappa_D^{0.25})$ steps. Thus an overall complexity becomes 
\begin{equation}
    T_{D^{-1/2}}=O\qty(\qty(\frac{\kappa_D^{0.625}}{\epsilon_D^{2.5}\epsilon_m^{0.5}}+\frac{\kappa_D^{0.5625}}{\epsilon_D^{2.25}\epsilon_e\epsilon_m^{0.75}})\log^2 N).
\end{equation}}
The last step in constructing the {symmetrized} transition matrix is multiplying inverse {square-root} degree matrix to $\hat{K}$, which
we perform in the exponential (Sec. \ref{subsec:invert}).
Therefore, we first need to exponentiate {$D^{-1/2}$} and $\hat{K}$ using a time of $O(t^2/(m^2\epsilon_e))$ and $4n^\prime$ copies each.
The time complexity for constructing $e^{iX_3(S)t}$ is therefore
{
\begin{equation*}
    T_S(t)=O\qty(T_{D^{-1/2}}\cdot\qty(\frac{t^{1.5}}{\epsilon_m^{0.5}\epsilon_e}+\frac{t}{\epsilon_e})+\frac{t^{1.5}}{\epsilon_m^{0.5}\epsilon_e}\log N).
\end{equation*}}


\subsection{Eigen-decomposition\label{subsec:eigendecompose}}

In this subsection, we find the eigenvalues and eigenvectors of $S$, 
using a second QPE algorithm (Sec. \ref{sec: QPE2}).
We take $t=O(1/\epsilon_S)$ to perform the QPE in time $O(\log N)$ with accuracy $\epsilon_S$~\cite{dervovic2018HHLprimer}.

The third and final QPE is required to transform $\ket{s_i}$ to the right eigenvector $\ket{v_i} = D^{-1/2}\ket{s_i}$ of the transition matrix $P = D^{-1}K$.
The complexity analysis proceeds analogously to the analysis in Sec~\ref{sec: complexity_S}: the error analysis of matrix inversion gives $t = O(\kappa_D^{0.5}/\epsilon_D)$. 
As the success probability of measuring $\ket{1}$ in the ancilla qubit is $\Omega(1/\kappa_D)$, the number of rounds necessary to obtain a successful result is $O(\kappa_D)$.
To accelerate the process, amplitude amplification is used to amplify the success probability in $O(\kappa_D^{0.5})$ steps. Incorporating the time required to prepare copies of $e^{iX_3(D)t}$, the time complexity of the conditional rotation is hence $O(T_D(\kappa_D^{0.5}/\epsilon_D)\cdot\kappa_D^{0.5}\log N/\epsilon_D)$.
Thus, the overall complexity of the eigen-decomposition of the transition matrix becomes
\begin{equation}
    O\qty(T_S\qty(\frac{1}{\epsilon_S})\cdot\frac{\log N}{\epsilon_S}+T_D\qty(\frac{\kappa_D^{0.5}}{\epsilon_D})\cdot\frac{\kappa_D^{0.5}\log N}{\epsilon_D}),
\end{equation}
which proves Theorem \ref{thm:main}.

We repeat the QPE and measure the eigenvalue register in the computation basis until we find sufficiently many copies of all eigenpairs for tomography. The process is not deterministic and we can only talk about the expected number of repetitions required.
The task of drawing items (with replacement) distributed according to a given probability distribution $\{p_i\}_{i=1}^N$ until one collects all $N$ distinct items is known as the coupon collector's problem, and an upper bound on the expected number of draws is known to be $(\max_i \{p_i\}/\min_i \{p_i\}) N\log N$ \cite{cCouponCollector}.
Once we repeat the QPE and measure the eigenvalue register in the computation basis until we obtain $O(N \log N/\epsilon_t)$ copies of each eigenpair, the pure-state tomography algorithm of  \cite{Kerenidis_interior} recovers a classical vector that is $\epsilon_t$-close in $\ell_2$-norm to the eigenvector. The above facts together with Theorem \ref{thm:main} immediately implies Lemma \ref{lemma: classical_coupon}.

Finally, in the case of uniformly random coupons i.e. $p_i = p_j$ for all $i,j\in\{1,\dots,N\}$, there exists a gate-efficient quantum algorithm that solves the coupon collector problem in time $O(N)$ \cite{qCouponCollector}, giving Lemma \ref{lemma: quantum_coupon}.

\end{document}